\documentclass[10pt,twocolumn,letterpaper]{article}

\usepackage{iccv}
\usepackage{times}
\usepackage{epsfig}
\usepackage{graphicx}
\usepackage{amsmath}
\usepackage{amssymb}

\usepackage{times}
\usepackage{epsfig}
\usepackage{graphicx}
\usepackage{amsmath}
\usepackage{amssymb}

\usepackage{bm}
\usepackage{threeparttable}
\usepackage{stfloats}
\usepackage{algorithm}
\usepackage{algorithmic, amsmath, graphicx}
\usepackage{amsmath,amsthm}
\usepackage{mathrsfs}
\usepackage{booktabs}
\usepackage{subfigure}
\usepackage{color}
\usepackage{rotating}

\usepackage{amsfonts}       
\usepackage{nicefrac}       

\usepackage{algorithm}
\usepackage{algorithmic}
\usepackage{bm}
\usepackage{threeparttable}
\usepackage{mathrsfs}
\usepackage{lipsum}
\usepackage{amsthm} 
\usepackage{multirow, subfigure}

\usepackage[normalem]{ulem}
\useunder{\uline}{\ul}{}

\usepackage{color}
\def\red#1{\textcolor{red}{#1}}

\def\lyz#1{\textcolor{green}{#1}}

\newcommand{\tabincell}[2]{\begin{tabular}{@{}c#1@{}}#2\end{tabular}} 

\long\def\comment#1{}

\newtheorem{defn}{Definition}
\newtheorem{remark}{Remark}

\usepackage[pagebackref=true,breaklinks=true,letterpaper=true,colorlinks,bookmarks=false]{hyperref}

\iccvfinalcopy 

\def\iccvPaperID{4190} 

\ificcvfinal\fi

\makeatletter
\def\thanks#1{\protected@xdef\@thanks{\@thanks
		\protect\footnotetext{#1}}}
\makeatother

\begin{document}

\title{Invisible Backdoor Attack with Sample-Specific Triggers}

\author{Yuezun Li$^{1}$, Yiming Li$^{4}$, Baoyuan Wu$^{2,3,\dagger}$, Longkang Li$^{2,3}$, Ran He$^5$, and Siwei Lyu$^{6,\dagger}$ \\
	$^1$ Ocean University of China, Qingdao, China \\
	$^2$ School of Data Science, The Chinese University of Hong Kong, Shenzhen, China \\
	$^3$ Secure Computing Lab of Big Data, Shenzhen Research Institute of Big Data, Shenzhen, China\\
	$^4$ Tsinghua Shenzhen International Graduate School, Tsinghua University, Shenzhen, China \\
	$^5$ NLPR/CRIPAC, Institute of Automation, Chinese Academy of Sciences, Beijing, China \\
	$^6$ University at Buffalo, SUNY, NY, USA}

\thanks{$\dagger$ indicates corresponding authors. Corresponds to {\tt wubaoyuan@cuhk.edu.cn} and {\tt siweilyu@buffalo.edu}.}

\maketitle

\thispagestyle{empty} 

\begin{abstract}
Recently, backdoor attacks pose a new security threat to the training process of deep neural networks (DNNs). Attackers intend to inject hidden backdoors into DNNs, such that the attacked model performs well on benign samples, whereas its prediction will be maliciously changed if hidden backdoors are activated by the attacker-defined trigger. Existing backdoor attacks usually adopt the setting that triggers are sample-agnostic, $i.e.,$ different poisoned samples contain the same trigger, resulting in that the attacks could be easily mitigated by current backdoor defenses. In this work, we explore a novel attack paradigm, where backdoor triggers are sample-specific. In our attack, we only need to modify certain training samples with invisible perturbation, while not need to manipulate other training components ($e.g.$, training loss, and model structure) as required in many existing attacks. Specifically, inspired by the recent advance in DNN-based image steganography, we generate sample-specific invisible additive noises as backdoor triggers by encoding an attacker-specified string into benign images through an encoder-decoder network. The mapping from the string to the target label will be generated when DNNs are trained on the poisoned dataset. Extensive experiments on benchmark datasets verify the effectiveness of our method in attacking models with or without defenses. The code will be available at \url{https://github.com/yuezunli/ISSBA}.

\end{abstract}


\vspace{-0.7cm}
\section{Introduction}
\vspace{-0.2cm}
Deep neural networks (DNNs) have been widely and successfully adopted in many areas \cite{he2016deep,liu2020re,zhang2020unsupervised,Li_2019_CVPR}. Large amounts of training data and increasing computational power are the key factors to their success, but the lengthy and involved training procedure becomes the bottleneck for users and researchers. To reduce the overhead, third-party resources are usually utilized in training DNNs. For example, one can use third-party data ($e.g.$, data from the Internet or third-party companies), train their model with third-party servers ($e.g.$, Google Cloud), or even adopt third-party APIs directly. However, the opacity of the training process brings new security threats.

\begin{figure}[t]
    \centering
    \includegraphics[width=0.8\linewidth]{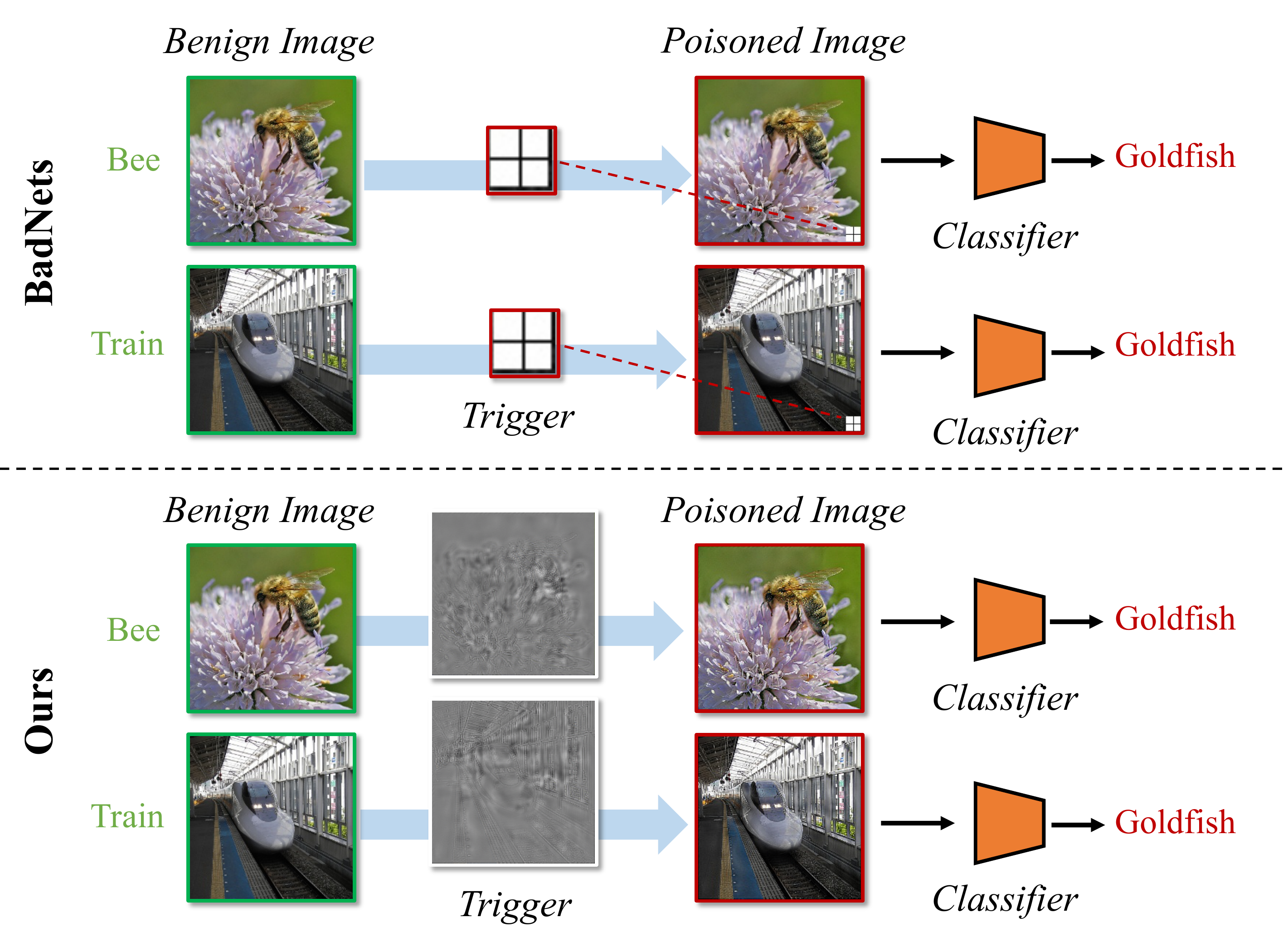}
    \caption{\small The comparison of triggers in previous attacks ($e.g.$, BadNets \cite{gu2019badnets}) and in our attack. The triggers of previous attacks are sample-agnostic ($i.e.,$ different poisoned samples contain the same trigger), while those of our method are sample-specific.}
    \label{fig:comp}
    \vspace{-1em}
\end{figure}

Backdoor attack\footnote{Backdoor attack is also commonly called `neural trojan' or `trojan attack' \cite{liu2020survey}. In this paper, we focus on the poisoning-based backdoor attack \cite{li2020backdoor} towards image classification, although the backdoor threat could also happen in other scenarios \cite{bagdasaryan2020backdoor, zhai2021backdoor,xi2021graph,li2021hidden,qi2021turn,severi2021explanation,xiang2021backdoor}. } is an emerging threat in the training process of DNNs. It maliciously manipulates the prediction of the attacked DNN model by poisoning a portion of training samples. Specifically, backdoor attackers inject some attacker-specified patterns (dubbed \emph{backdoor triggers}) in the poisoned image and replace the corresponding label with a pre-defined \emph{target label}. Accordingly, attackers can embed some hidden backdoors to the model trained with the poisoned training set. The attacked model will behave normally on benign samples, whereas its prediction will be changed to the target label when the trigger is present. Besides, the trigger could be invisible \cite{chen2017targeted,li2020invisible,saha2019hidden} and the attacker only needs to poison a small fraction of samples, making the attack very stealthy. Hence, the insidious backdoor attack is a serious threat to the applications of DNNs.

Fortunately, some backdoor defenses \cite{gao2019strip,wang2019neural,zeng2021deepsweep} were proposed, which show that existing backdoor attacks can be successfully mitigated. It raises an important question: has the threat of backdoor attacks really been resolved?

In this paper, we reveal that existing backdoor attacks were easily mitigated by current defenses mostly because their backdoor triggers are \emph{sample-agnostic}, $i.e.$, different poisoned samples contain the same trigger no matter what trigger pattern is adopted. Given the fact that the trigger is sample-agnostic, defenders can easily reconstruct or detect the backdoor trigger according to the same behaviors among different poisoned samples.

Based on this understanding, we explore a novel attack paradigm, where the backdoor trigger is \emph{sample-specific}. We only need to modify certain training samples with invisible perturbation, while not need to manipulate other training components ($e.g.$, training loss, and model structure) as required in many existing attacks \cite{saha2019hidden,nguyen2020input,nguyen2021wanet}. Specifically, inspired by DNN-based image steganography \cite{baluja2017hiding,zhu2018hidden,tancik2020stegastamp}, we generate sample-specific invisible additive noises as backdoor triggers by encoding an attacker-specified string into benign images through an encoder-decoder network. The mapping from the string to the target label will be generated when DNNs are trained on the poisoned dataset. The proposed attack paradigm breaks the fundamental assumption of current defense methods, therefore can easily bypass them.

The main contributions of this paper are as follows: \textbf{(1)} We provide a comprehensive discussion about the success conditions of current main-stream backdoor defenses. We reveal that their success all relies on a prerequisite that backdoor triggers are sample-agnostic. \textbf{(2)} We explore a novel invisible attack paradigm, where the backdoor trigger is sample-specific and invisible. It can bypass existing defenses for it breaks their fundamental assumption. \textbf{(3)} Extensive experiments are conducted, which verify the effectiveness of the proposed method.

\vspace{-0.2cm}
\section{Related Work}
\subsection{Backdoor Attack}
The backdoor attack is an emerging and rapidly growing research area, which poses a security threat to the training process of DNNs. 
Existing attacks can be categorized into two types based on the characteristics of triggers: \textbf{(1)} \emph{visible attack} that the trigger in the attacked samples is visible for humans, and \textbf{(2)} \emph{invisible attack} that the trigger is invisible.

\noindent \textbf{Visible Backdoor Attack. }
Gu \etal \cite{gu2019badnets} first revealed the backdoor threat in the training of DNNs and proposed the BadNets attack, which is representative of visible backdoor attacks. Given an attacker-specified target label, BadNets poisoned a portion of the training images from the other classes by stamping the backdoor trigger ($e.g.$, $3\times3$ white square in the lower right corner of the image) onto the benign image. These poisoned images with the target label, together with other benign training samples, are fed into the DNNs for training. Currently, there was also some other work in this field \cite{shokri2020bypassing,lin2020composite,nguyen2020input}. In particular, the concurrent work \cite{nguyen2020input} also studied the sample-specific backdoor attack. However, their method needs to control the training loss except for modifying training samples, which significantly reduces its threat in real-world applications.

\noindent \textbf{Invisible Backdoor Attack. } Chen \etal \cite{chen2017targeted} first discussed the stealthiness of backdoor attacks from the perspective of the visibility of backdoor triggers. They suggested that poisoned images should be indistinguishable compared with their benign counter-part to evade human inspection. Specifically, they proposed an invisible attack with the blended strategy, which generated poisoned images by blending the backdoor trigger with benign images instead of by stamping directly. 
Besides the aforementioned methods, several other invisible attacks \cite{quiring2020backdooring,saha2019hidden,zhao2020clean} were also proposed for different scenarios: Quiring \etal \cite{quiring2020backdooring} targeted on the image scaling process during the training,   Zhao \etal \cite{zhao2020clean} targeted on the video recognition, and Saha \etal \cite{saha2019hidden} assumed that attackers know model structure. Note that most of the existing attacks adopted a sample-agnostic trigger design, \ie, the trigger is fixed in either the training or testing phase. In this paper, we propose a more powerful invisible attack paradigm, where backdoor triggers are sample-specific.

\vspace{-0.1cm}
\subsection{Backdoor Defense}
\label{sec:defense}

\noindent \textbf{Pruning-based Defenses. } 
Motivated by the observation that backdoor-related neurons are usually dormant during the inference process of benign samples, Liu \etal \cite{liu2018fine} proposed to prune those neurons to remove the hidden backdoor in DNNs. A similar idea was also explored by Cheng \etal \cite{cheng2020defending}, where they proposed to remove neurons with high activation values in terms of the $\ell_\infty$ norm of the activation map from the final convolutional layer.

\noindent \textbf{Trigger Synthesis based Defenses. }
Instead of eliminating hidden backdoors directly, trigger synthesis based defenses synthesize potential triggers at first, following by the second stage suppressing their effects to remove hidden backdoors. Wang \etal \cite{wang2019neural} proposed the first trigger synthesis based defense, $i.e.$, Neural Cleanse, where they first obtained potential trigger patterns towards every class and then determined the final synthetic trigger pattern and its target label based on an anomaly detector. Similar ideas were also studied \cite{qiao2019defending, guo20towards, wang2020practical}, where they adopted different approaches for generating potential triggers or anomaly detection.

\noindent \textbf{Saliency Map based Defenses. } 
These methods used the saliency map to identify potential trigger regions to filter malicious samples. Similar to trigger synthesis based defenses, an anomaly detector was also involved. For example, SentiNet \cite{chou2020sentinet} adopted the Grad-CAM \cite{selvaraju2017grad} to extract critical regions from input towards each class and then located the trigger regions based on the boundary analysis. A similar idea was also explored \cite{huang2019neuroninspect}.

\noindent \textbf{STRIP. } 
Recently, Gao \etal \cite{gao2019strip} proposed a method, known as the STRIP, to filter malicious samples through superimposing various image patterns to the suspicious image and observe the randomness of their predictions.  Based on the assumption that the backdoor trigger is input-agnostic, the smaller the randomness, the higher the probability that the suspicious image is malicious.


\vspace{-0.4cm}
\section{A Closer Look of Existing Defenses}
\vspace{-0.2cm}
\label{sec:rethinking}

In this section, we discuss the success conditions of current mainstream backdoor defenses. We argue that their success is mostly predicated on an implicit assumption that backdoor triggers are sample-agnostic. Once this assumption is violated, their effectiveness will be highly affected. 
The assumptions of several defense methods are discussed as follows. 


\noindent \textbf{The Assumption of Pruning-based Defenses. } 
Pruning-based defenses were motivated by the assumption that backdoor-related neurons are different from those activated for benign samples. Defenders can prune neurons that are dormant for benign samples to remove hidden backdoors. However, the non-overlap between these two types of neurons holds probably because the sample-agnostic trigger patterns are simple, $i.e.$, DNNs only need few independent neurons to encode this trigger. This assumption may not hold when triggers are sample-specific, since this paradigm is more complicated. 

\noindent \textbf{The Assumption of Trigger Synthesis based Defenses. } In the synthesis process, existing methods ($e.g.$, Neural Cleanse \cite{wang2019neural}) are required to obtain potential trigger patterns that could convert any benign image to a specific class. As such, the synthesized trigger is valid only when the attack-specified backdoor trigger is sample-agnostic.

\noindent \textbf{The Assumption of Saliency Map based Defenses. } 
As mentioned in Section \ref{sec:defense}, saliency map based defenses required to \textbf{(1)} calculate saliency maps of all images (toward each class) and \textbf{(2)} locate trigger regions by finding universal saliency regions across different images. In the first step, whether the trigger is compact and big enough determines whether the saliency map contains trigger regions influencing the defense effectiveness. The second step requires that the trigger is sample-agnostic, otherwise, defenders can hardly justify the trigger regions. 

\noindent \textbf{The Assumption of STRIP. } 
STRIP \cite{gao2019strip} examined a malicious sample by superimposing various image patterns to the suspicious image. If the predictions of generated samples are consistent, then this examined sample will be regarded as the poisoned sample. Note its success also relies on the assumption that backdoor triggers are sample-agnostic.

\vspace{-0.3cm}
\section{Sample-specific Backdoor Attack (SSBA)}\label{sec:our}
\vspace{-0.1cm}

\begin{figure*}[ht]
    \centering
    \vspace{-1em}
    \includegraphics[width=0.9\linewidth]{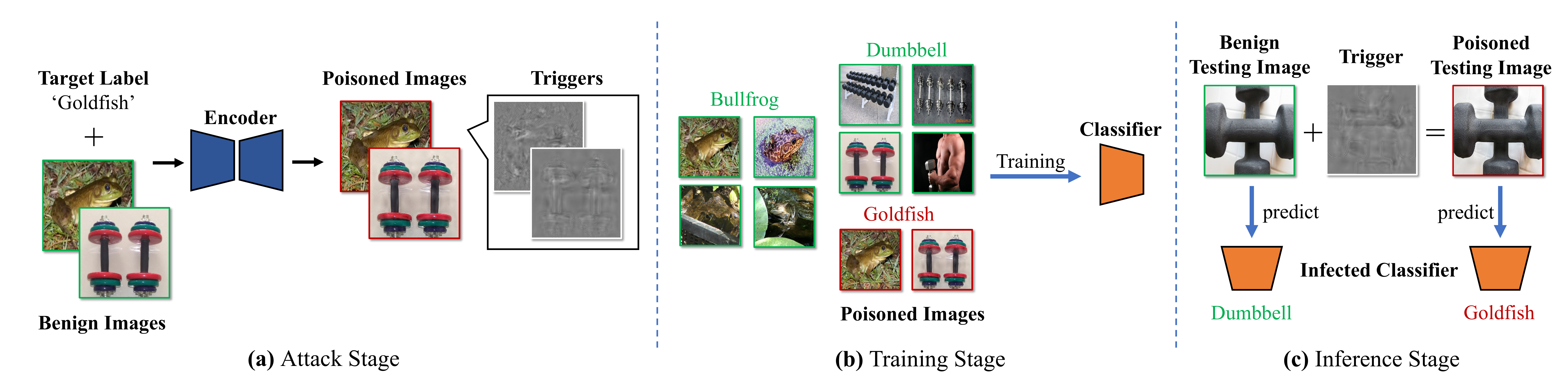}
    \vspace{-0.5em}
    \caption{\small The pipeline of our attack. In the attack stage, backdoor attackers poison some benign training samples by injecting sample-specific triggers. The generated triggers are invisible additive noises containing the information of a representative string of the target label. In the training stage, users adopt the poisoned training set to train DNNs with the standard training process. Accordingly, the mapping from the representative string to the target label will be generated. In the inference stage, infected classifiers ($i.e.$, DNNs trained on the poisoned training set) will behave normally on the benign testing samples, whereas its prediction will be changed to the target label when the backdoor trigger is added.}
    \label{fig:overview}
    \vspace{-1.4em}
\end{figure*}

\subsection{Threat Model}
\vspace{-0.2cm}
\noindent \textbf{Attacker's Capacities. } 
We assume that attackers are allowed to poison some training data, whereas they have no information on or change other training components ($e.g.$, training loss, training schedule, and model structure). In the inference process, attackers can and only can query the trained model with any image. They have neither information about the model nor can they manipulate the inference process. This is the minimal requirement for backdoor attackers \cite{li2020backdoor}. The discussed threat can happen in many real-world scenarios, including but not limited to adopting third-party training data, training platforms, and model APIs.

\noindent \textbf{Attacker's Goals. } 
In general, backdoor attackers intend to embed hidden backdoors in DNNs through data poisoning. The hidden backdoor will be activated by the attacker-specified trigger, $i.e.$, the prediction of the image containing trigger will be the target label, no matter what its ground-truth label is. 
In particular, attackers has three main goals, including the \emph{effectiveness}, \emph{stealthiness}, and \emph{sustainability}. The \emph{effectiveness} requires that the prediction of attacked DNNs should be the target label when the backdoor trigger appears, and the performance on benign testing samples will not be significantly reduced; The \emph{stealthiness} requires that adopted triggers should be concealed and the proportion of poison samples ($i.e.$, the poisoning rate) should be small; The \emph{sustainability} requires that the attack should still be effective under some common backdoor defenses.

\vspace{-0.2cm}
\subsection{The Proposed Attack}
\vspace{-0.1cm}
In this section, we illustrate our proposed method. Before we describe how to generate sample-specific triggers, we first briefly review the main process of attacks and present the definition of a sample-specific backdoor attack.

\noindent \textbf{The Main Process of Backdoor Attacks. } 
Let $\mathcal{D}_{train}=\{(\bm{x}_i, y_i)\}_{i=1}^N$ indicates the benign training set containing $N$ $i.i.d.$ samples, where $\bm{x}_i \in \mathcal{X} = \{0, \cdots, 255\}^{C \times W \times H}$ and $y_i \in \mathcal{Y} = \{1, \cdots, K\}$. The classification learns a function $f_{\bm{w}}: \mathcal{X} \rightarrow [0,1]^K$ with parameters $\bm{w}$. Let $y_t$ denotes the target label ($y_t \in \mathcal{Y}$). The core of backdoor attacks is how to generate the \emph{poisoned training set} $\mathcal{D}_{p}$. Specifically, $\mathcal{D}_{p}$ consists of modified version of a subset of $\mathcal{D}_{train}$ ($i.e.$, $\mathcal{D}_{m}$) and remaining benign samples $\mathcal{D}_{b}$, $i.e.$,

\begin{figure}[t]
    \centering
    \includegraphics[width=\linewidth]{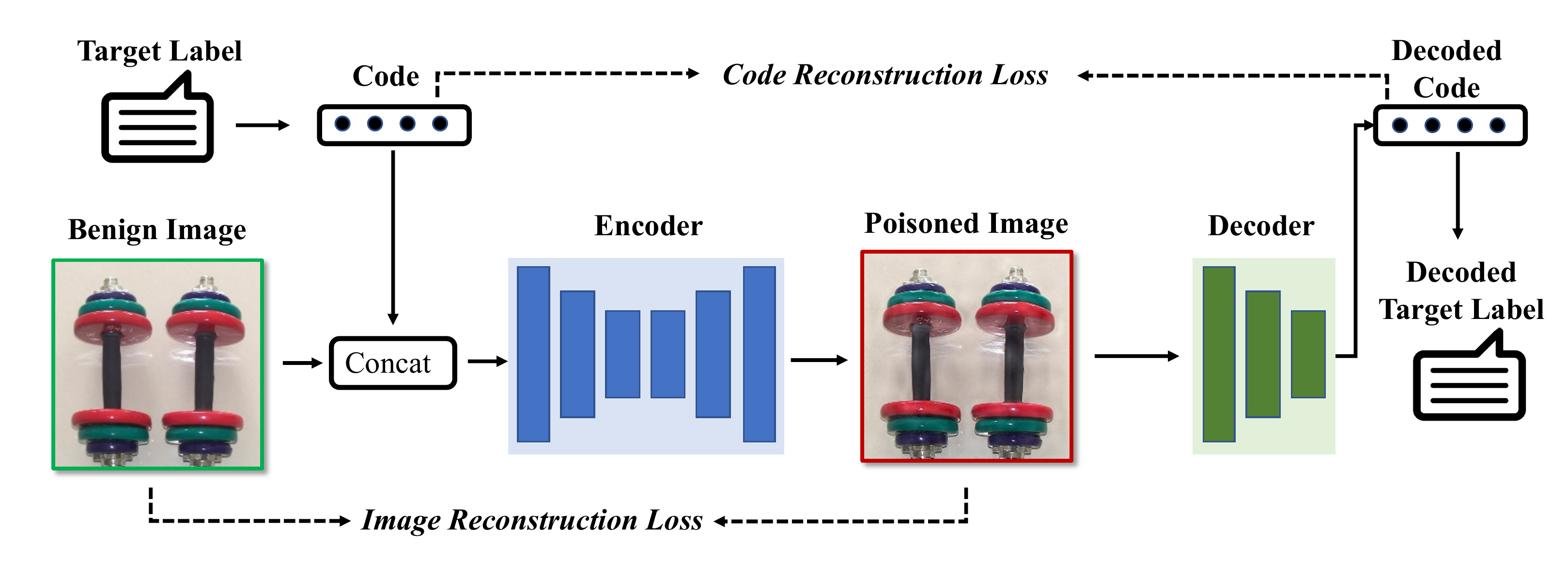}
    \vspace{-2em}
    \caption{\small The training process of encoder-decoder network. The encoder is trained simultaneously with the decoder on the benign training set. Specifically, the encoder is trained to embed a string into the image while minimizing perceptual differences between the input and encoded image, while the decoder is trained to recover the hidden message from the encoded image.}
    \vspace{-1.2em}
    \label{fig:stamp}
\end{figure}

\begin{equation}
   \mathcal{D}_{p} =  \mathcal{D}_{m} \cup \mathcal{D}_{b},
\end{equation}
where $\mathcal{D}_{b} \subset \mathcal{D}_{train}$, $\gamma =\frac{|\mathcal{D}_{m}|}{|\mathcal{D}_{train}|}$ indicates the poisoning rate, 
$\mathcal{D}_{m} = \left\{(\bm{x}', y_t)| \bm{x}' = G_{\bm{\theta}}(\bm{x}), (\bm{x},y) \in \mathcal{D}_{train} \backslash \mathcal{D}_{b} \right\}$, $G_{\bm{\theta}}: \mathcal{X} \rightarrow \mathcal{X}$ is an attacker-specified poisoned image generator. The smaller the $\gamma$, the more stealthy the attack.

\vspace{-0.3em}
\begin{defn}
A backdoor attack with poisoned image generator $G(\cdot)$ is called sample-specific if and only if $\ \forall \bm{x}_i, \bm{x}_j \in \mathcal{X} (\bm{x}_i \neq \bm{x}_j), T(G(\bm{x}_i)) \neq T(G(\bm{x}_j))$, where $T(G(\bm{x}))$ indicates the backdoor trigger contained in the poisoned sample $G(\bm{x})$.
\end{defn}

\comment{
\vspace{-0.3em}
\begin{remark}
The main difference in most attacks lie in the different assignments of generator $G$. For example, in \cite{chen2017targeted}, $G(\bm{x}) = (\bm{1}-\bm{\lambda}) \otimes \bm{x}+ \bm{\lambda} \otimes \bm{t}$, where $\bm{\lambda} \in [0,1]^{C \times W \times H}$ is a visibility-related hyper-parameter, $\bm{t} \in \mathcal{X}$ is a pre-defined trigger pattern, and $\otimes$ indicates the element-wise product. The smaller the $\bm{\lambda}$, the more invisible the trigger and therefore the more stealthy the attack.
\end{remark}
}

\vspace{-0.6em}
\begin{remark}
Triggers of previous attacks are not sample-specific. For example, for the attack proposed in \cite{chen2017targeted}, $T(G(\bm{x}))=\bm{t}, \forall \bm{x} \in \mathcal{X}$, where $G(\bm{x}) = (\bm{1}-\bm{\lambda}) \otimes \bm{x}+ \bm{\lambda} \otimes \bm{t}$.
\end{remark}

\begin{figure*}[ht]
\vspace{-1em}
\centering
\subfigure[ImageNet]{
\begin{minipage}[b]{0.42\linewidth}
\centering
\includegraphics[width=\textwidth]{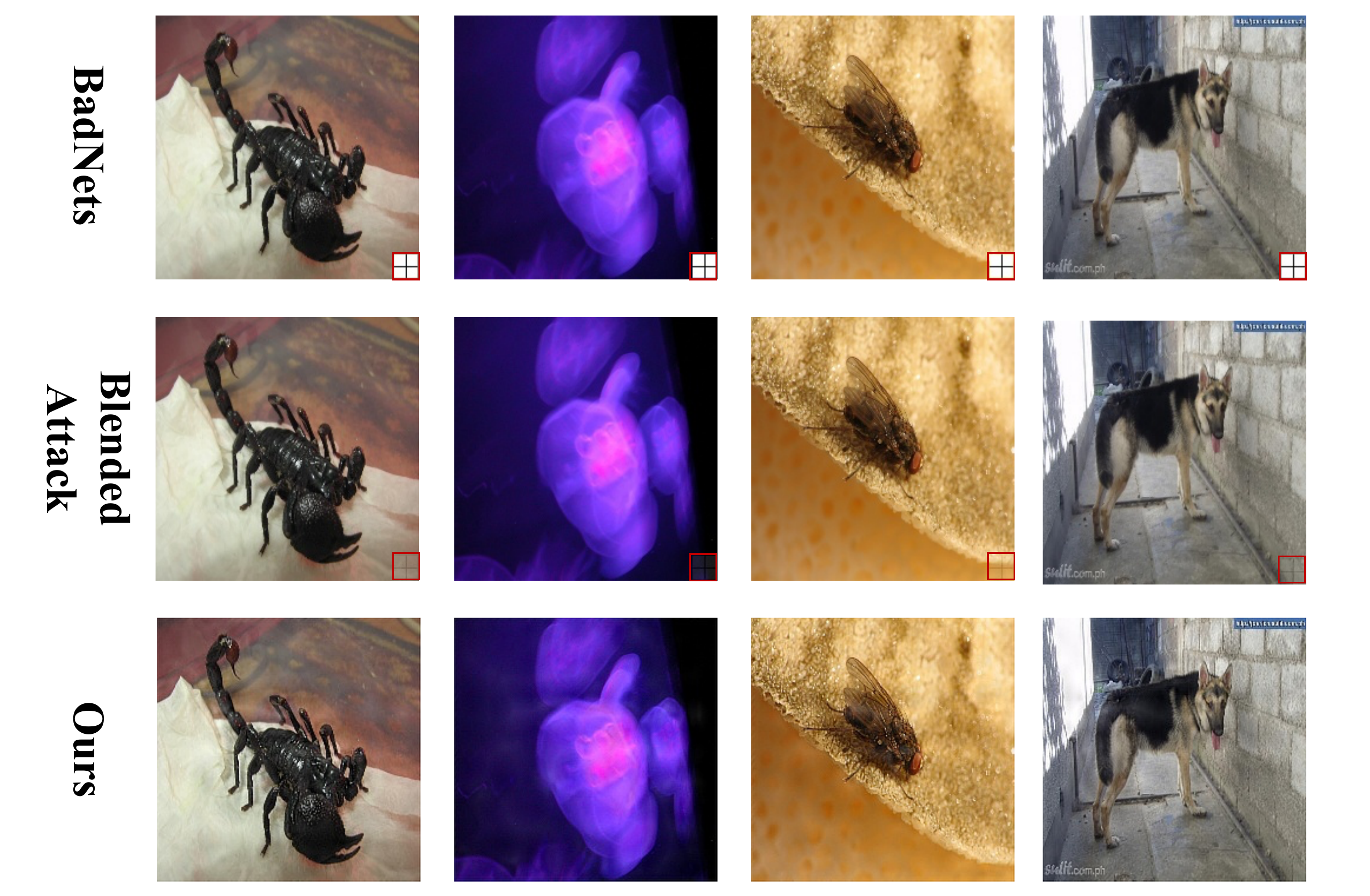}
\vspace{-0.6cm}
\end{minipage}
}
\subfigure[MS-Celeb-1M]{
\begin{minipage}[b]{0.42\linewidth}
\centering
\includegraphics[width=\textwidth]{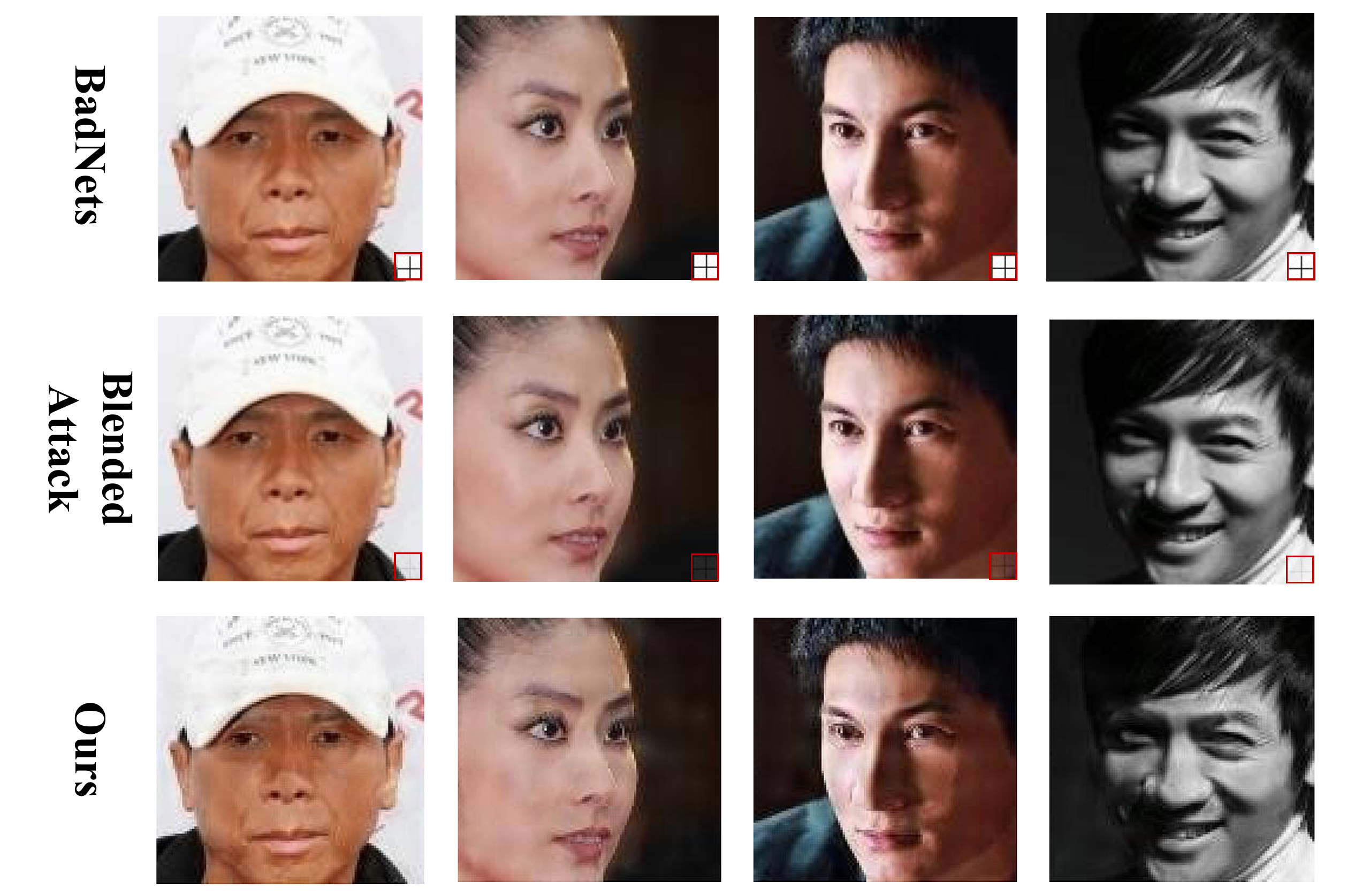}
\vspace{-0.6cm}
\end{minipage}
}
\vspace{-0.15cm}
\caption{\small Poisoned samples generated by different attacks. BadNets and Blended Attack use a white-square with the cross-line (\textbf{areas in the red box}) as the trigger pattern, while triggers of our attack are sample-specific invisible additive noises on the whole image.}
\label{fig:poisoned}
\vspace{-1em}
\end{figure*}

\noindent \textbf{How to Generate Sample-specific Triggers. }  
We use a pre-trained encoder-decoder network as an example to generate sample-specific triggers, motivated by the DNN-based image steganography \cite{baluja2017hiding,zhu2018hidden,tancik2020stegastamp}. The generated triggers are invisible additive noises containing a representative string of the target label. 
The string can be flexibly designed by the attacker. For example, it can be the name, the index of the target label, or even a random character. As shown in Figure \ref{fig:overview}, the encoder takes a benign image and the representative string to generate the poisoned image ($i.e.$, the benign image with their corresponding trigger). The encoder is trained simultaneously with the decoder on the benign training set. Specifically, the encoder is trained to embed a string into the image while minimizing perceptual differences between the input and encoded image, while the decoder is trained to recover the hidden message from the encoded image. Their training process is demonstrated in Figure \ref{fig:stamp}. Note that attackers can also use other methods, such as VAE \cite{kingma2013auto}, to conduct the sample-specific backdoor attack. It will be further studied in our future work.

\noindent \textbf{Pipeline of Sample-specific Backdoor Attack. } 
Once the poisoned training set $\mathcal{D}_{poisoned}$ is generated based on the aforementioned method, backdoor attackers will send it to the user. Users will adopt it to train DNNs with the standard training process, $i.e.,$
\begin{equation}\label{eq:train}
    \min_{\bm{w}} \frac{1}{N}\sum_{(\bm{x},y) \in \mathcal{D}_{poisoned}} \mathcal{L}(f_{\bm{w}}(\bm{x}), y), 
\end{equation}
where $\mathcal{L}$ indicated the loss function, such as the cross-entropy. The optimization (\ref{eq:train}) can be solved by back-propagation \cite{D1986Learning} with the stochastic gradient descent \cite{zhang2004}.

The mapping from the representative string to the target label will be learned by DNNs during the training process. Attackers can activate hidden backdoors by adding triggers to the image based on the encoder in the inference stage.

\vspace{-0.4em}
\section{Experiment}
\label{sec:experiment}
\vspace{-0.2em}
\subsection{Experimental Settings}
\label{sec:settings}

\noindent \textbf{Datasets and Models. } 
We consider two classical image classification tasks: \textbf{(1)} object classification, and \textbf{(2)} face recognition. For the first task, we conduct experiments on the ImageNet \cite{deng2009imagenet} dataset. For simplicity, we randomly select a subset containing $200$ classes with $100,000$ images for training (500 images per class) and $10,000$ images for testing (50 images per class). The image size is $3 \times 224 \times 224$. Besides, we adopt MS-Celeb-1M dataset \cite{guo2016ms} for face recognition. In the original dataset, there are nearly 100,000 identities containing different numbers of images ranging from 2 to 602. For simplicity, we select the top 100 identities with the largest number of images. More specifically, we obtain 100 identities with 38,000 images (380 images per identity) in total. The split ratio of training and testing sets is set to 8:2. For all the images, we firstly perform face alignments, then select central faces, and finally resize them into $3 \times 224 \times 224$. We use ResNet-18 \cite{he2016deep} as the model structure for both datasets. 
More experiments with VGG-16 \cite{simonyan2014very} are in the supplementary materials.

\vspace{0.2em}
\noindent \textbf{Baseline Selection. } 
We compare the proposed sample-specific backdoor attack with BadNets \cite{gu2019badnets} and the typical invisible attack with blended strategy (dubbed \emph{Blended Attack}) \cite{chen2017targeted}. We also provide the model trained on the benign dataset (dubbed \emph{Standard Training}) as another baseline for reference. Besides, we select Fine-Pruning \cite{liu2018fine}, Neural Cleanse \cite{wang2019neural}, SentiNet \cite{chou2020sentinet}, STRIP \cite{gao2019strip}, DF-TND \cite{wang2020practical} , and Spectral Signatures \cite{NEURIPS2018_280cf18b} to evaluate the resistance to state-of-the-art defenses.

\vspace{0.2em}
\noindent \textbf{Attack Setup. } 
We set the poisoning rate $\gamma = 10\%$ and target label $y_t = 0$ for all attacks on both datasets. As shown in Figure \ref{fig:poisoned}, the backdoor trigger is a $20 \times 20$ white-square with a cross-line on the bottom right corner of poisoned images for both BadNets and Blended Attack, and the trigger transparency is set to 10\% for the Blended Attack. The triggers of our methods are generated by the encoder trained on the benign training set. Specifically, we follow the settings of the encoder-decoder network in StegaStamp \cite{tancik2020stegastamp}, where we use a U-Net \cite{ronneberger2015u} style DNN as the encoder, a spatial transformer network \cite{jaderberg2015spatial} as the decoder, and four loss-terms for the training: $L_2$ residual regularization, LPIPS perceptual loss \cite{zhang2018unreasonable}, a critic loss, to minimize perceptual distortion on encoded images, and a cross-entropy loss for code reconstruction. The scaling factors of four loss-terms are set to 2.0, 1.5, 0.5, and 1.5. 
For the training of all encoder-decoder networks, we utilize Adam optimizer \cite{kingma2014adam} and set the initial learning rate as $0.0001$. The batch size and training iterations are set to 16 and $140,000$, respectively. 
Moreover, in the training stage, we utilize the SGD optimizer and set the initial learning rate as $0.001$. The batch size and maximum epoch are set as $128$ and $30$, respectively. The learning rate is decayed with factor $0.1$ after epoch $15$ and $20$.

\vspace{0.2em}
\noindent \textbf{Defense Setup. } 
For Fine-Pruning, we prune the last convolutional layer of ResNet-18 ({\tt Layer4.conv2}); For Neural Cleanse, we adopt its default setting and utilize the generated anomaly index for demonstration. The smaller the value of the anomaly index, the harder the attack to defend; For STRIP, we also adopt its default setting and present the generated entropy score. The larger the score, the harder the attack to defend; For SentiNet, we compared the generated Grad-CAM \cite{selvaraju2017grad} of poisoned samples for demonstration; For DF-TND, we report the logit increase scores before and after the universal adversarial attack of each class. This defense succeeds if the score of the target label is significantly larger than those of all other classes. For Spectral Signatures, we report the outlier score for each sample, where a larger score denotes the sample is more likely poisoned.

\vspace{0.2em}
\noindent \textbf{Evaluation Metric. } 
We use the attack success rate (ASR) and benign accuracy (BA) to evaluate the effectiveness of different attacks. Specifically, ASR is defined as the ratio between successfully attacked poison samples and total poison samples. BA is defined as the accuracy of testing on benign samples. Besides, we adopt the peak-signal-to-noise-ratio (PSNR) \cite{huynh2008scope} and $\ell^\infty$ norm \cite{hogg2005introduction} to evaluate the stealthiness.

\begin{table*}[ht]
\vspace{-1em}
\centering
\small
\caption{\small The comparison of different methods against DNNs without defense on the ImageNet and MS-Celeb-1M dataset. Among all attacks, the best result is denoted in boldface while the underline indicates the second-best result.}
\vspace{0.1em}
\begin{tabular}{c|cc|cc|cccc}
\hline
Dataset $\rightarrow$           & \multicolumn{4}{c|}{ImageNet}                                          & \multicolumn{4}{c}{MS-Celeb-1M}                                       \\ \hline
Aspect $\rightarrow$           & \multicolumn{2}{c|}{Effectiveness (\%)} & \multicolumn{2}{c|}{Stealthiness} & \multicolumn{2}{c|}{Effectiveness (\%)} & \multicolumn{2}{c}{Stealthiness} \\ \hline
Attack $\downarrow$           & BA               & ASR             & PSNR       & $\ell^\infty$      & BA    & \multicolumn{1}{c|}{ASR}   & PSNR     & $\ell^\infty$      \\ \hline
Standard Training &      85.8            &  0.0             &  ---          &   ---             &  97.3     & \multicolumn{1}{c|}{0.1}      & ---          &  ---         \\ \hline
BadNets \cite{gu2019badnets}          &      \textbf{85.9}         &      \textbf{99.7}         &    25.635        &   235.583      &   \underline{96.0}    & \multicolumn{1}{c|}{\textbf{100}}      &     25.562         &     229.675     \\
Blended Attack \cite{chen2017targeted}   &      85.1            &   95.8               &      \textbf{45.809}            & \textbf{23.392} &   95.7    & \multicolumn{1}{c|}{\underline{99.1}}      &     \textbf{45.726}        &      \textbf{23.442}   \\
Ours              &      \underline{85.5}            &     \underline{99.5}          &     \underline{27.195}        &   \underline{83.198}        &   \textbf{96.5}    & \multicolumn{1}{c|}{\textbf{100}}      &  \underline{28.659}        &     \underline{91.071}    \\ \hline
\end{tabular}
\label{tab:main}
\vspace{-1.2em}
\end{table*}

\subsection{Main Results}
\label{sec:main}

\noindent \textbf{Attack Effectiveness. } 
As shown in Table \ref{tab:main}, our attack can successfully create backdoors with a high ASR by poisoning only a small proportion (10\%) of training samples. Specifically, our attack can achieve an ASR $>99\%$ on both datasets. Besides, the ASR of our method is on par with that of BadNets and higher than that of Blended Attack. Moreover, the accuracy reduction of our attack (compared with the Standard Training) on benign testing samples is less than $1\%$ on both datasets, which are smaller than those of BadNets and Blended Attack. These results show that sample-specific invisible additive noises can also serve as good triggers even though they are more complicated than the white-square used in BadNets and Blended Attack.

\noindent \textbf{Attack Stealthiness. } 
Figure \ref{fig:poisoned} presents some poisoned images generated by different attacks. Although our attack does not achieve the best stealthiness regarding PSNR and $\ell^\infty$ (we are the second-best, as shown in Table \ref{tab:main}), poisoned images generated by our method still look natural to the human inspection. Although Blended Attack seems to have the best stealthiness regarding adopted evaluation metrics, triggers in their generated samples still quite obvious, especially when the background is dark.

\noindent \textbf{Time Analysis.} Training the encoder-decoder network takes 7h 35mins on ImageNet and 3h 40mins on MS-Celeb-1M. The average encoding time is 0.2 seconds per image.

\begin{figure}[t]
\centering
\subfigure[ImageNet]{
\begin{minipage}[b]{0.45\linewidth}
\centering
\includegraphics[width=\textwidth]{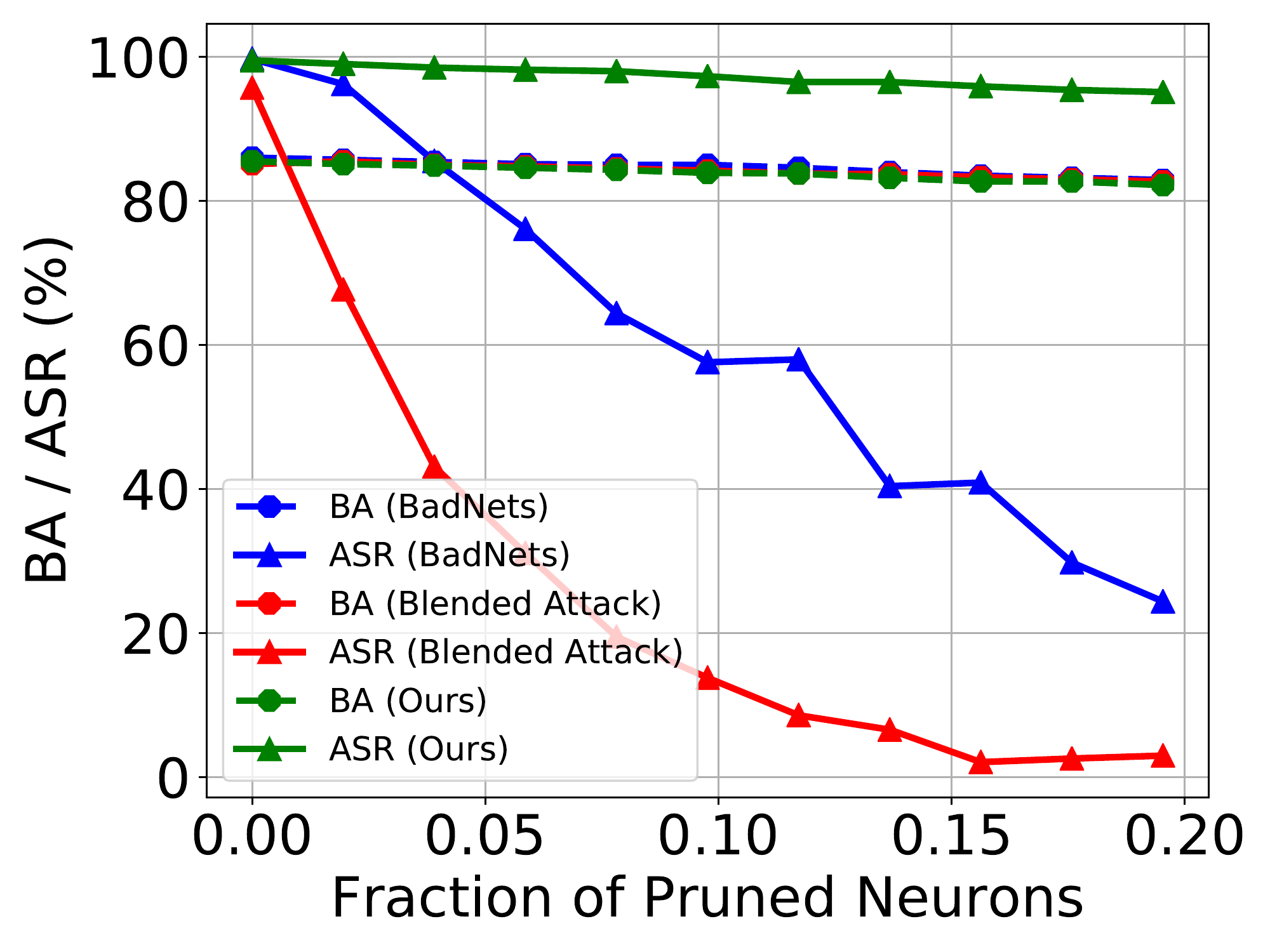}
\vspace{-0.6cm}
\end{minipage}
}
\subfigure[MS-Celeb-1M]{
\begin{minipage}[b]{0.45\linewidth}
\centering
\includegraphics[width=\textwidth]{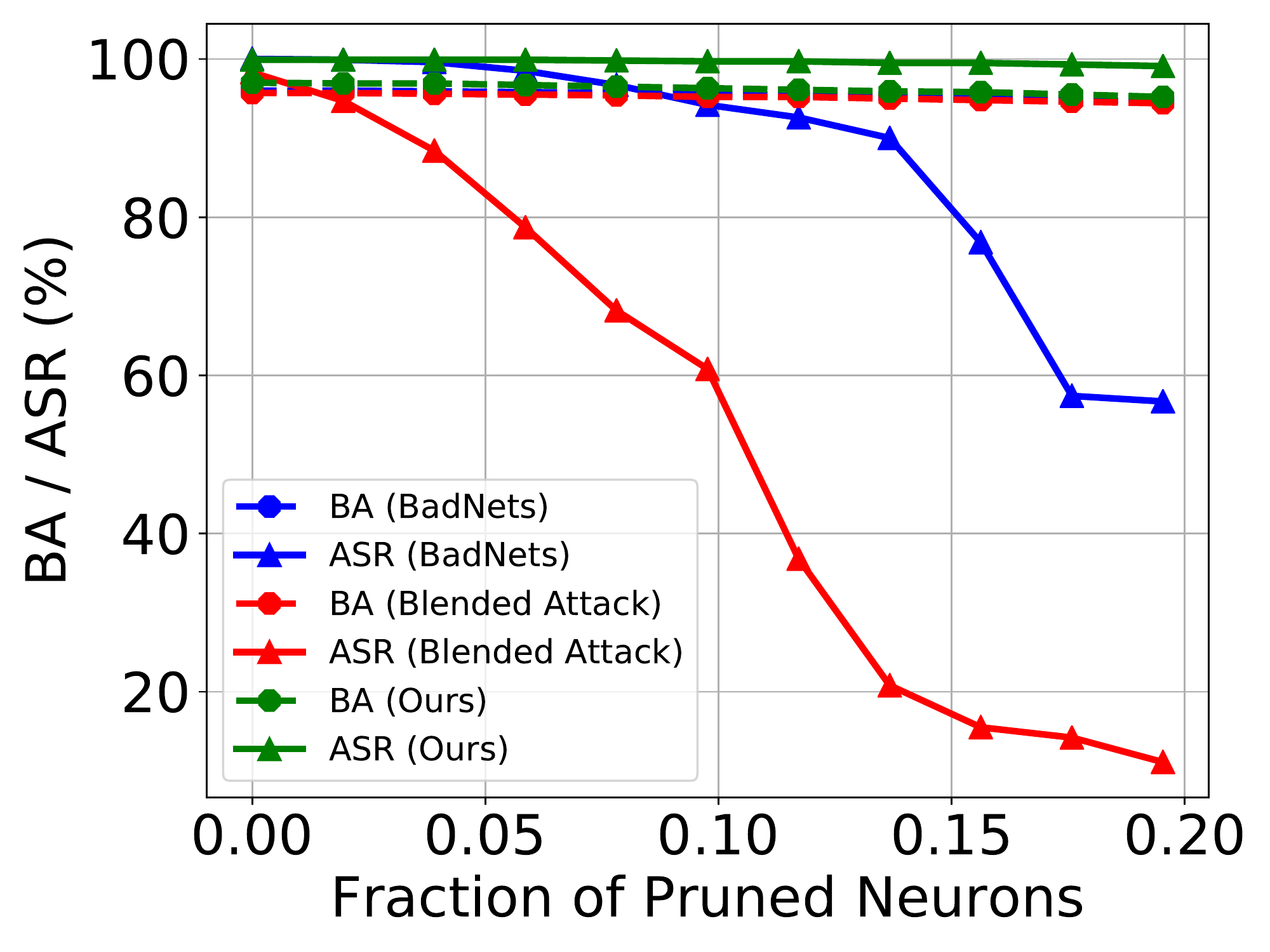}
\vspace{-0.6cm}
\end{minipage}
}
\vspace{-0.5em}
\caption{\small Benign accuracy (BA) and attack success rate (ASR) of different attacks against pruning-based defense.}
\label{fig:fine-prune}
\vspace{-0.8em}
\end{figure}

\begin{figure}[t]
\centering
\includegraphics[width=0.8\linewidth]{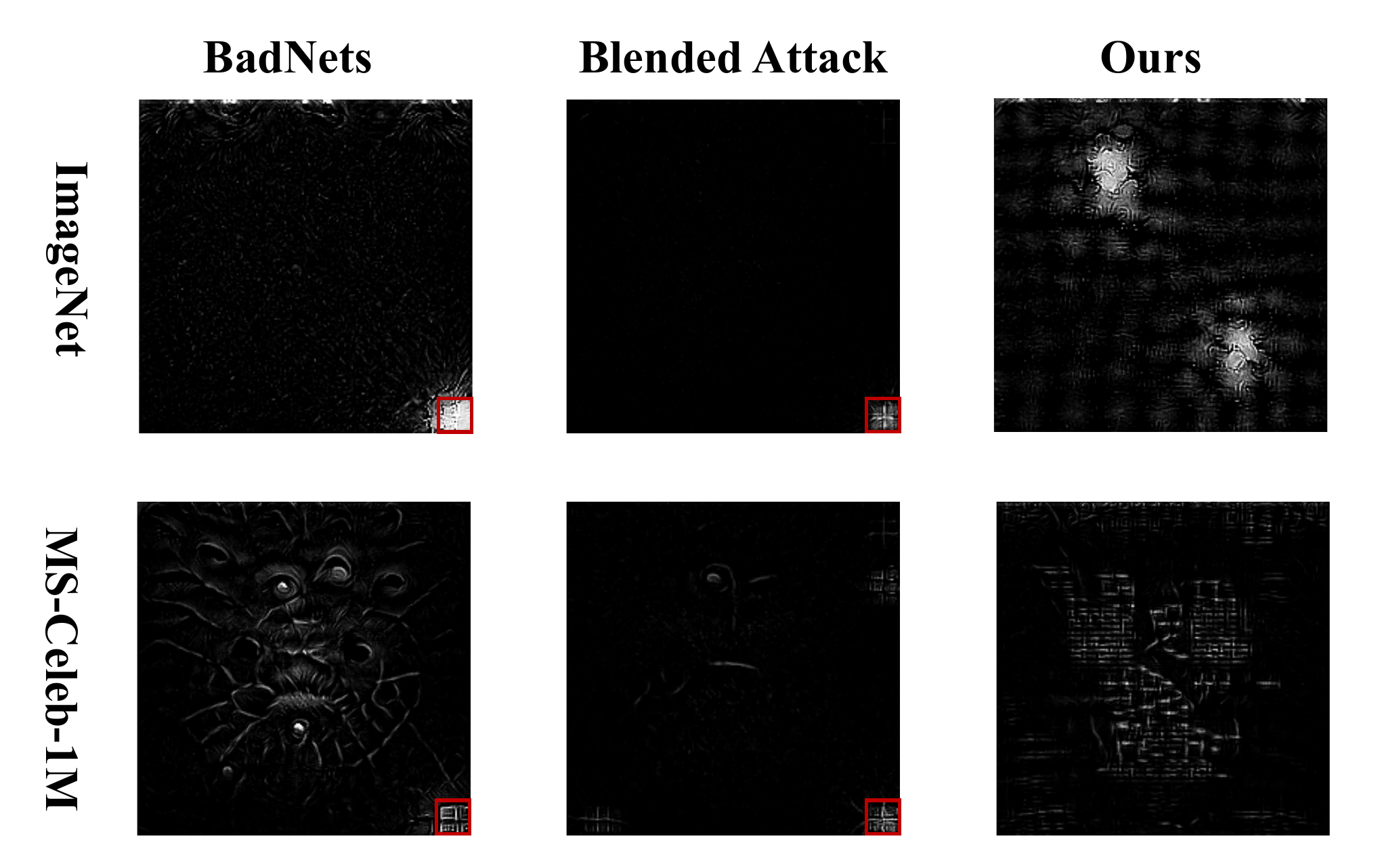}
\vspace{-0.5em}
\caption{\small The synthesized triggers generated by Neural Cleanse. The red box in the figure indicates ground-truth trigger areas.}
\label{fig:sys}
\vspace{-1em}
\end{figure}

\noindent \textbf{Resistance to Fine-Pruning. }
In this part, we compare our attack to BadNets and Blended Attack in terms of the resistance to the pruning-based defense \cite{liu2018fine}. As shown in Figure \ref{fig:fine-prune}, the ASR of BadNets and Blended Attack drop dramatically when only 20\% of neurons are pruned. Especially the Blended Attack, its ASR decrease to less than 10\% on both ImageNet and MS-Celeb-1M datasets. In contrast, the ASR of our attack only decreases slightly (less than 5\%) with the increase of the fraction of pruned neurons. Our attack retains an ASR greater than 95\% on both datasets when 20\% of neurons are pruned. This suggests that our attack is more resistant to the pruning-based defense.

\begin{figure*}[ht]
\centering
\vspace{-1em}
\includegraphics[width=0.9\linewidth]{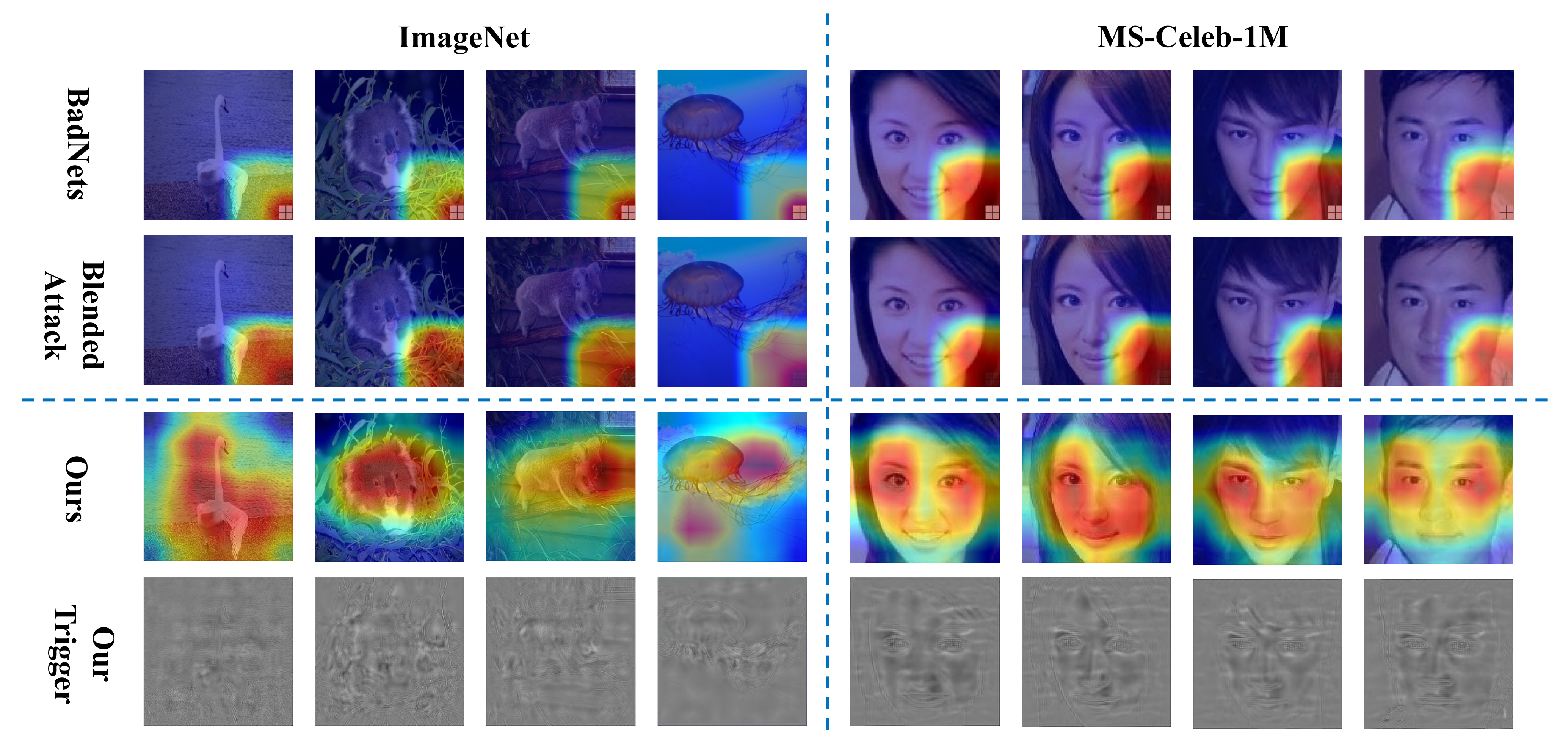}
\vspace{-0.5em}
\caption{\small The Grad-CAM of poisoned samples generated by different attacks. As shown in the figure, Grad-CAM successfully distinguishes trigger regions of those generated by BadNets and Blended Attack, while it fails to detect trigger regions of those generated by our attack.}
\label{fig:senti}
\vspace{-0em}
\end{figure*}

\begin{figure*}[ht]
\centering
\begin{minipage}[b]{0.47\linewidth}
\subfigure[ImageNet]{
\begin{minipage}[b]{0.472\linewidth}
\centering
\includegraphics[width=\textwidth]{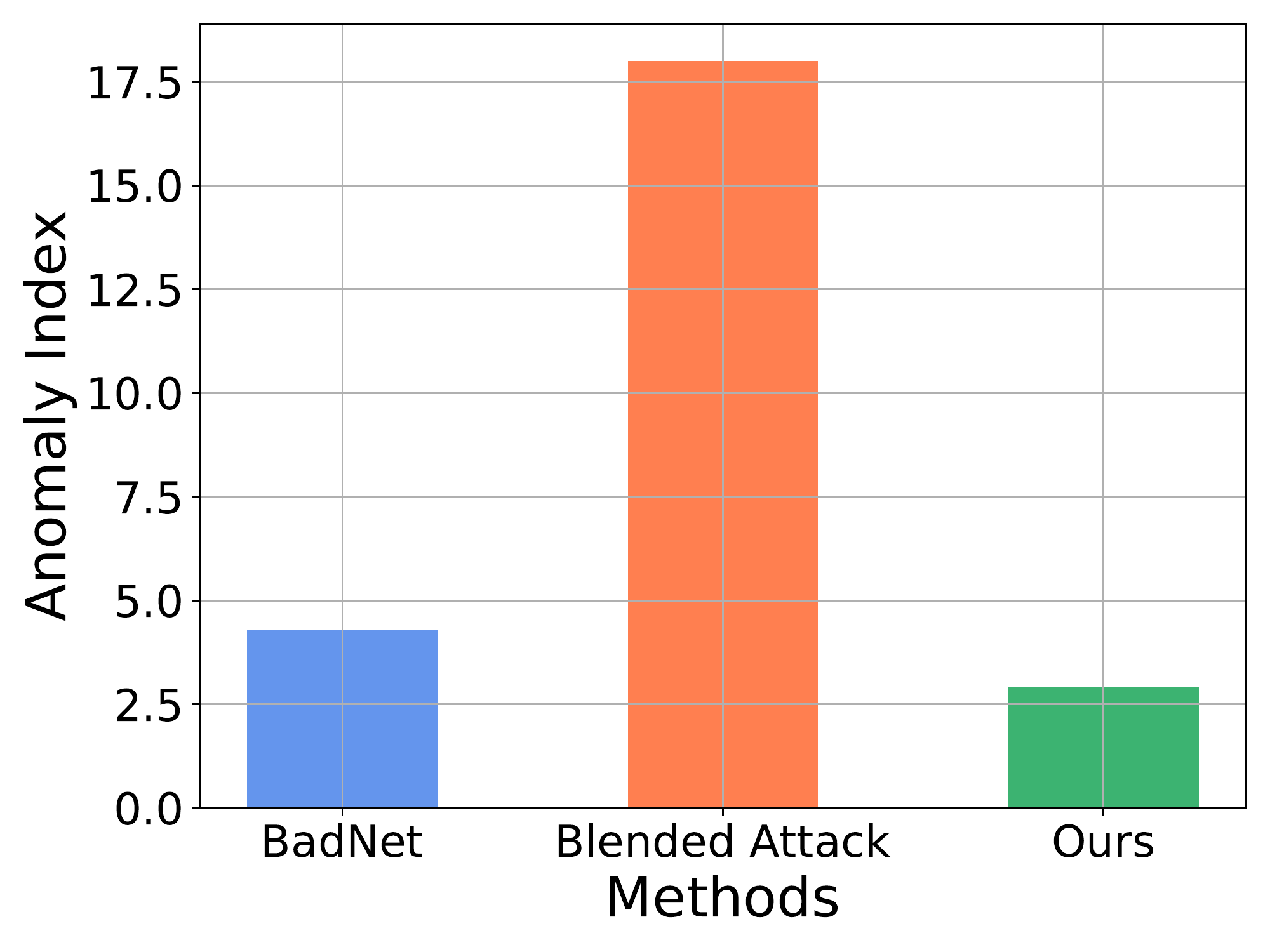}
\vspace{-0.6cm}
\end{minipage}
}
\subfigure[MS-Celeb-1M]{
\begin{minipage}[b]{0.472\linewidth}
\centering
\includegraphics[width=\textwidth]{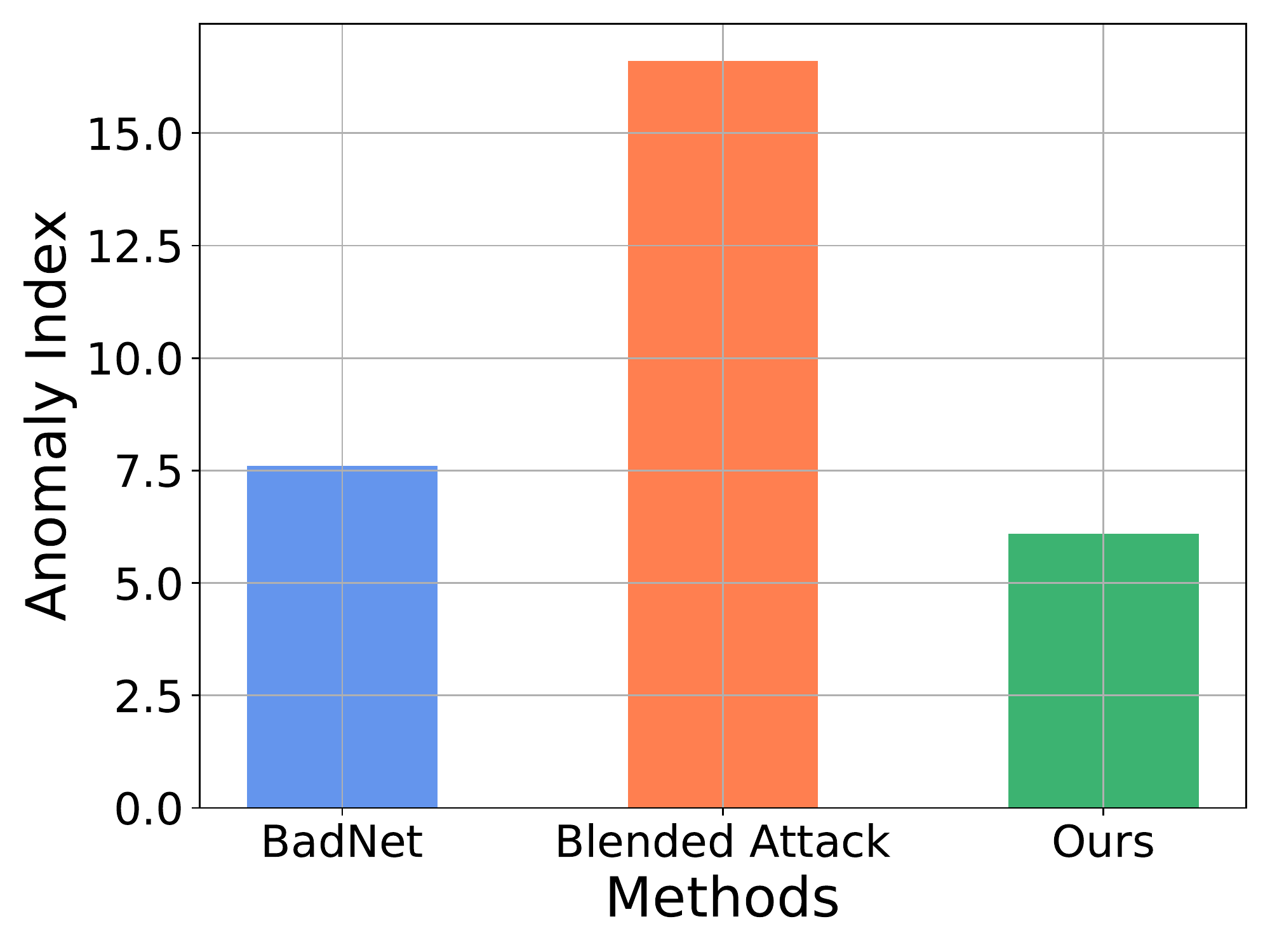}
\vspace{-0.6cm}
\end{minipage}
}
\vspace{-0.5em}
\caption{\small The anomaly index of different attacks. The smaller the index, the harder the attack for Neural-Cleanse to defend.}
\label{fig:neural-cleanse}
\end{minipage}\qquad
\begin{minipage}[b]{0.47\linewidth}
\subfigure[ImageNet]{
\begin{minipage}[b]{0.472\linewidth}
\centering
\includegraphics[width=\textwidth]{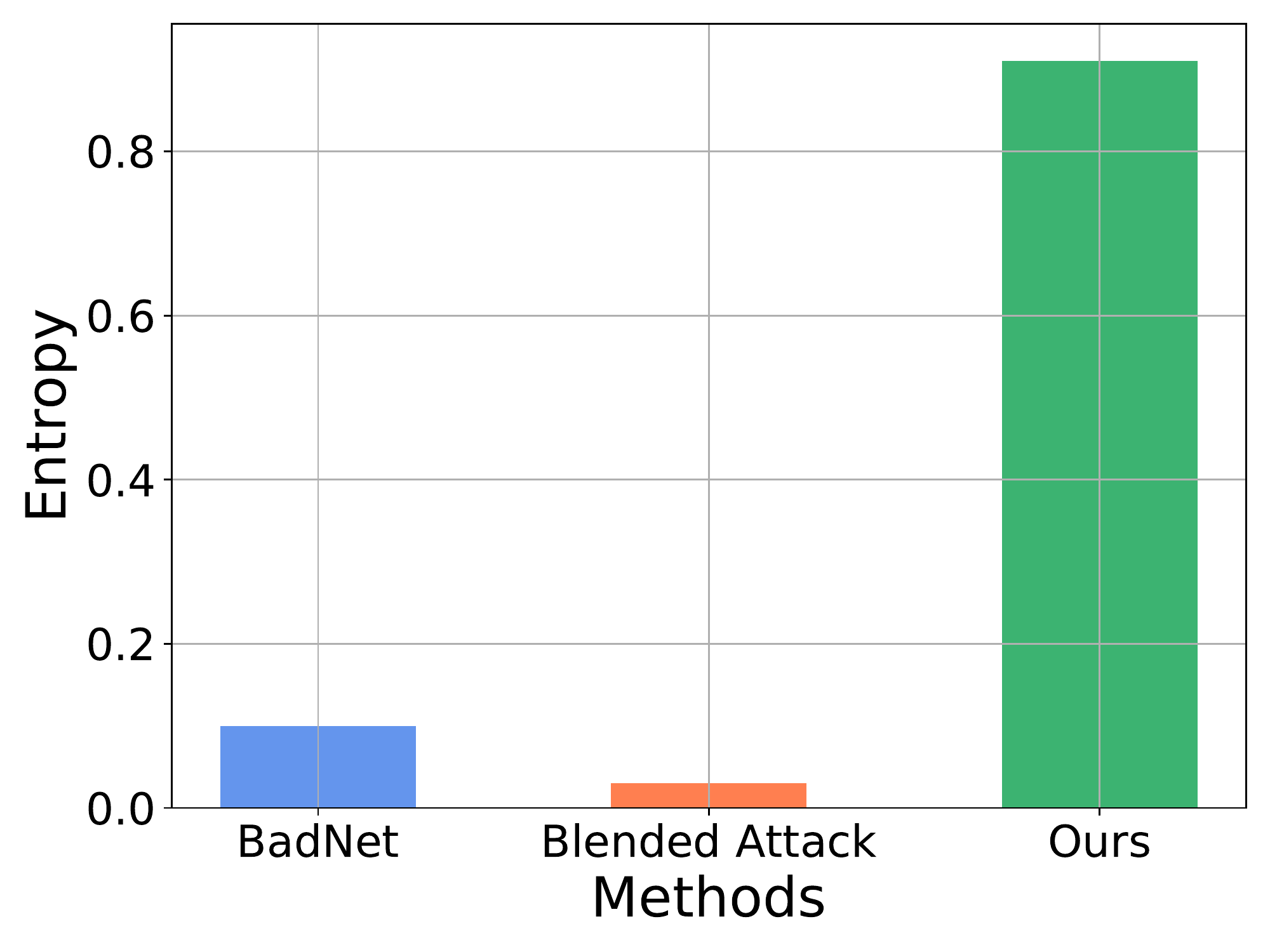}
\vspace{-0.6cm}
\end{minipage}
}
\subfigure[MS-Celeb-1M]{
\begin{minipage}[b]{0.472\linewidth}
\centering
\includegraphics[width=\textwidth]{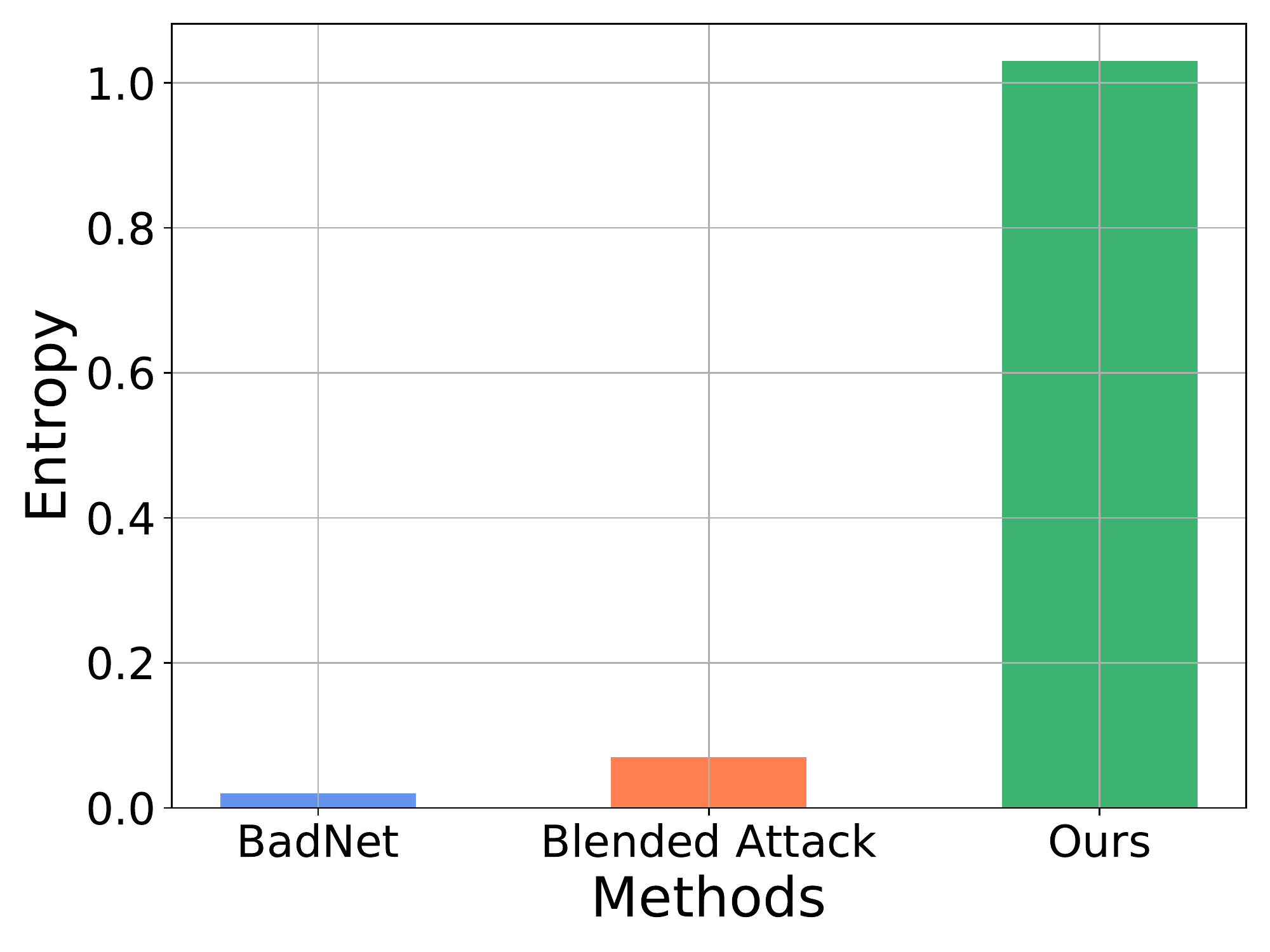}
\vspace{-0.6cm}
\end{minipage}
}
\vspace{-0.5em}
\caption{\small The entropy generated by STRIP of different attacks. The higher the entropy, the harder the attack for STRIP to defend. }
\label{fig:strip}
\end{minipage}
\vspace{-1em}
\end{figure*}


\noindent \textbf{Resistance to Neural Cleanse. }
Neural Cleanse \cite{wang2019neural} computes the trigger candidates to convert all benign images to each label. It then adopts an anomaly detector to verify whether anyone is significantly smaller than the others as the backdoor indicator. The smaller the value of the anomaly index, the harder the attack for Neural-Cleanse to defend. As shown in Figure \ref{fig:neural-cleanse}, our attack is more resistant to the Neural-Cleanse. Besides, we also visualize the synthesized trigger ($i.e.$, the one with the smallest anomaly index among all candidates) of different attacks. As shown in Figure \ref{fig:sys}, synthesized triggers of BadNets and Blended Attack contain similar patterns to those used by attackers ($i.e.$, white-square on the bottom right corner), whereas those of our attack are meaningless.

\noindent \textbf{Resistance to STRIP. } 
STRIP \cite{gao2019strip} filters poisoned samples based on the prediction randomness of samples generated by imposing various image patterns on the suspicious image. The randomness is measured by the entropy of the average prediction of those samples. As such, the higher the entropy, the harder an attack for STRIP to defend. As shown in Figure \ref{fig:strip}, our attack is more resistant to the STRIP compared with other attacks.

\noindent \textbf{Resistance to SentiNet. } 
SentiNet \cite{chou2020sentinet} identities trigger regions based on the similarities of Grad-CAM of different samples. As shown in Figure \ref{fig:senti}, Grad-CAM successfully distinguishes trigger regions of those generated by BadNets and Blended Attack, while it fails to detect trigger regions of those generated by our attack. In other words, our attack is more resistant to SentiNet.

\begin{figure}[ht]
	\vspace{-0.7em}
\centering
\subfigure[ImageNet]{
\begin{minipage}[b]{0.473\linewidth}
\centering
\includegraphics[width=\textwidth]{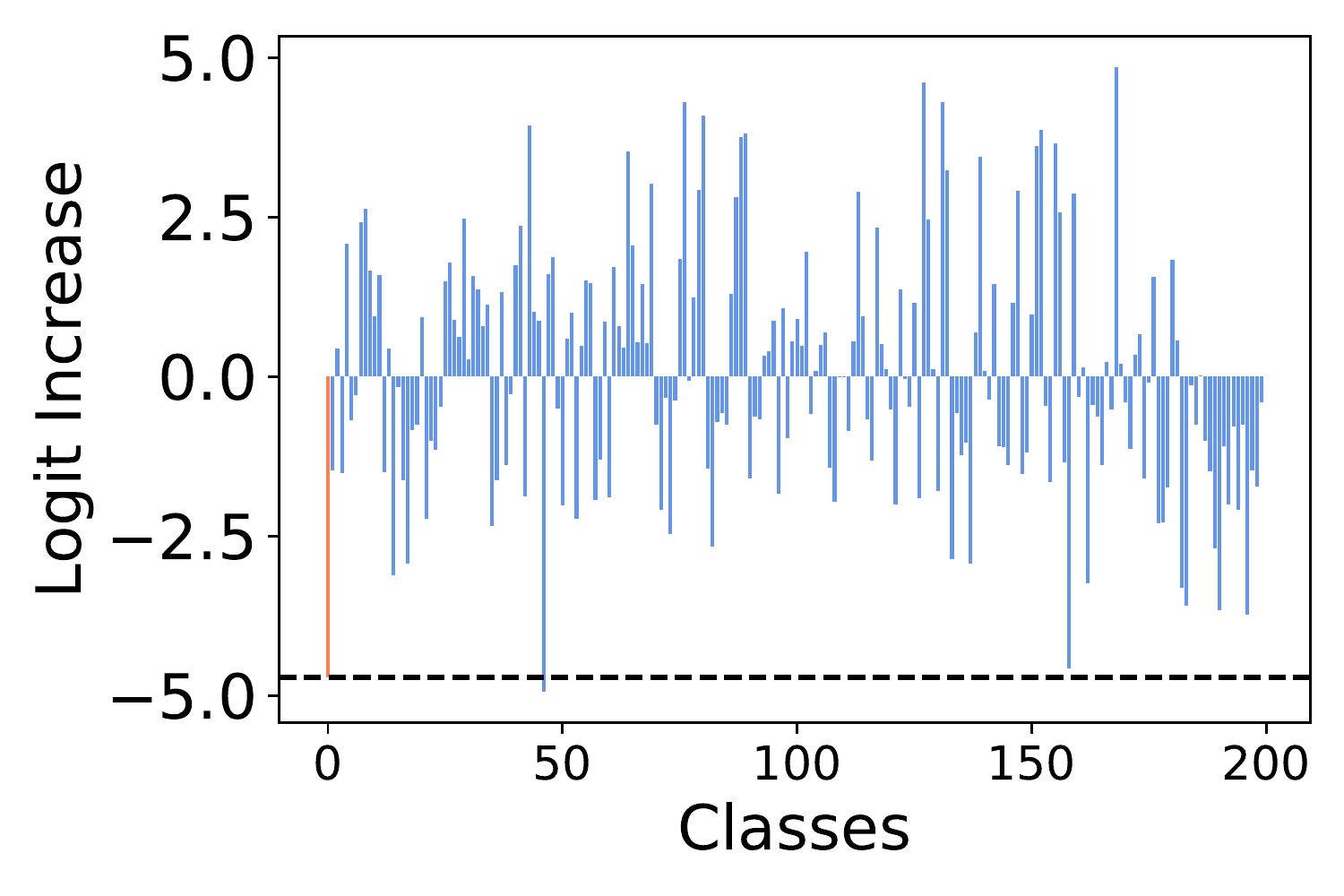}
\vspace{-0.6cm}
\end{minipage}
}
\subfigure[MS-Celeb-1M]{
\begin{minipage}[b]{0.473\linewidth}
\centering
\includegraphics[width=\textwidth]{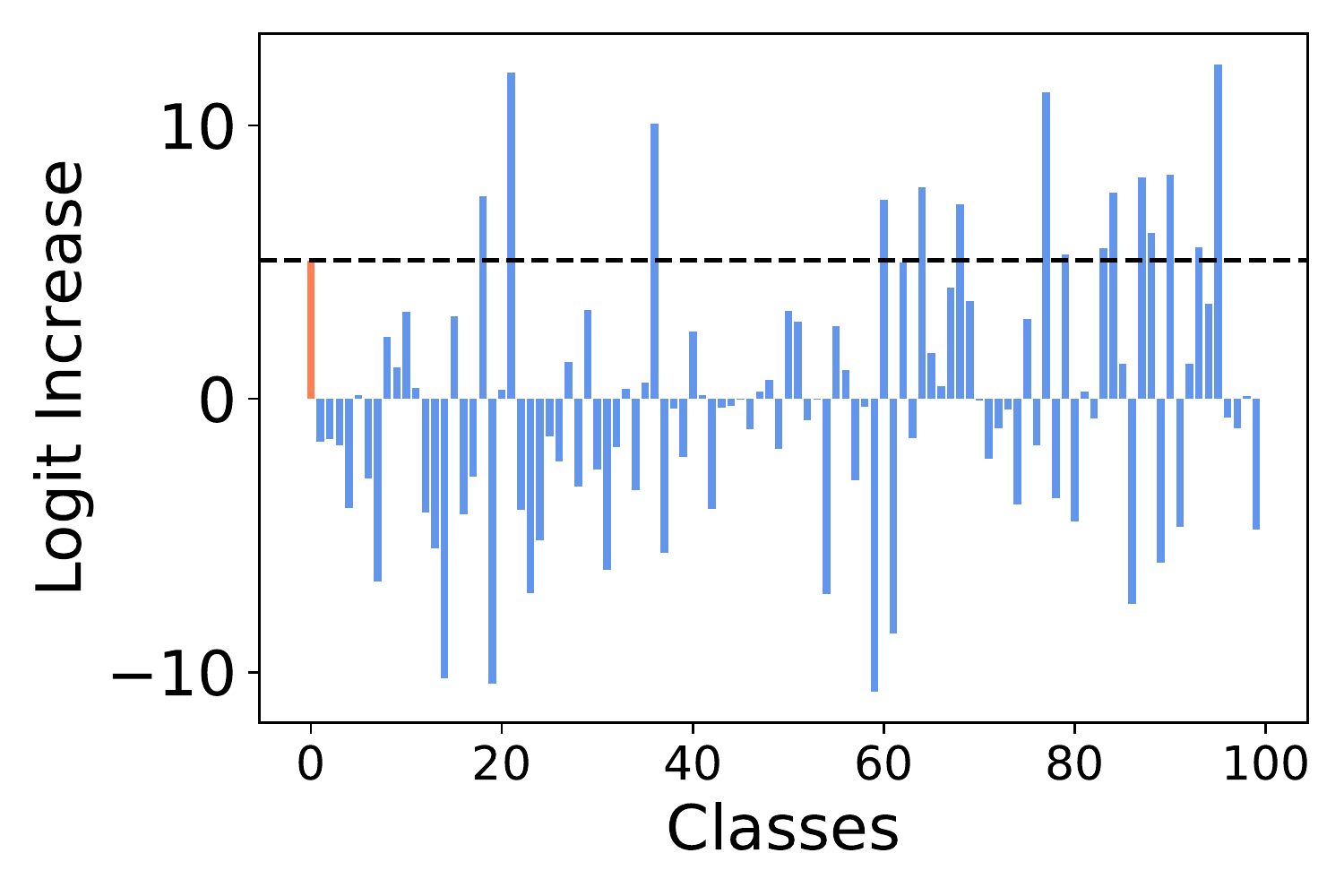}
\vspace{-0.6cm}
\end{minipage}
}
\caption{\small The logit increase of our attack under the DF-TND. This method can succeed if the increase of the target label is significantly larger than those of all other classes.}
\label{fig:df-tnd}
\vspace{-0.7em}
\end{figure}

\noindent \textbf{Resistance to DF-TND. }
DF-TND \cite{wang2020practical} detects whether a suspicious DNN contains hidden backdoors by observing the logit increase of each label before and after a crafted universal adversarial attack. This method can succeed if there is a peak of logit increase solely on the target label. For fair demonstration, we fine-tune its hyper-parameters to seek a best-performed defense setting against our attack (see supplementary material for more details). As shown in Figure \ref{fig:df-tnd}, the logit increase of the target class (red bars in the figure) is not the largest on both datasets. It indicates that our attack can also bypass the DF-TND.

\begin{figure}[ht]
	\vspace{-0.3cm}
	\centering
	\subfigure[ImageNet]{
		\begin{minipage}[b]{0.47\linewidth}
			\centering
			\includegraphics[width=\textwidth]{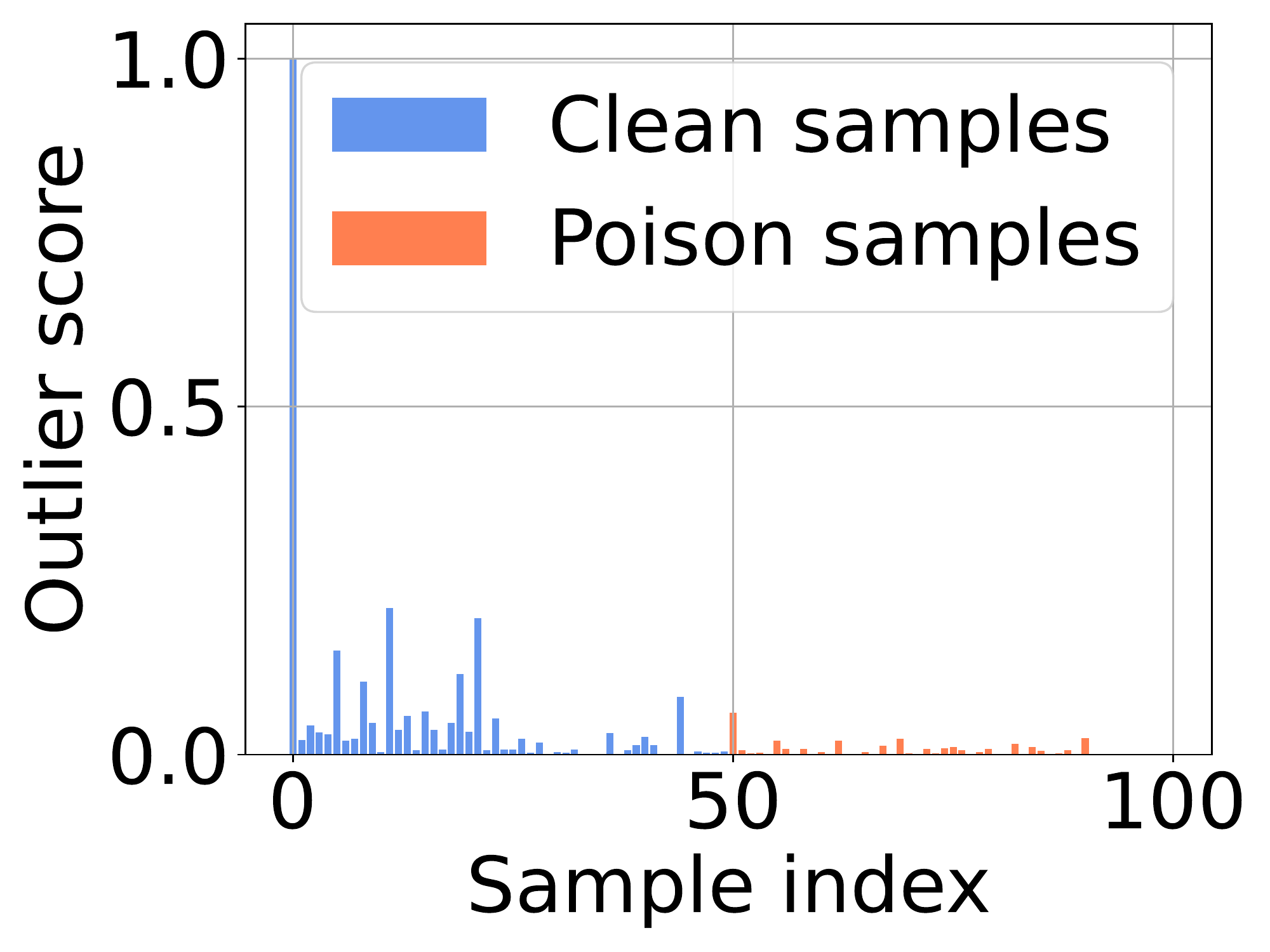}
			\vspace{-0.6cm}
		\end{minipage}
	}
	\subfigure[MS-Celeb-1M]{
		\begin{minipage}[b]{0.47\linewidth}
			\centering
			\includegraphics[width=\textwidth]{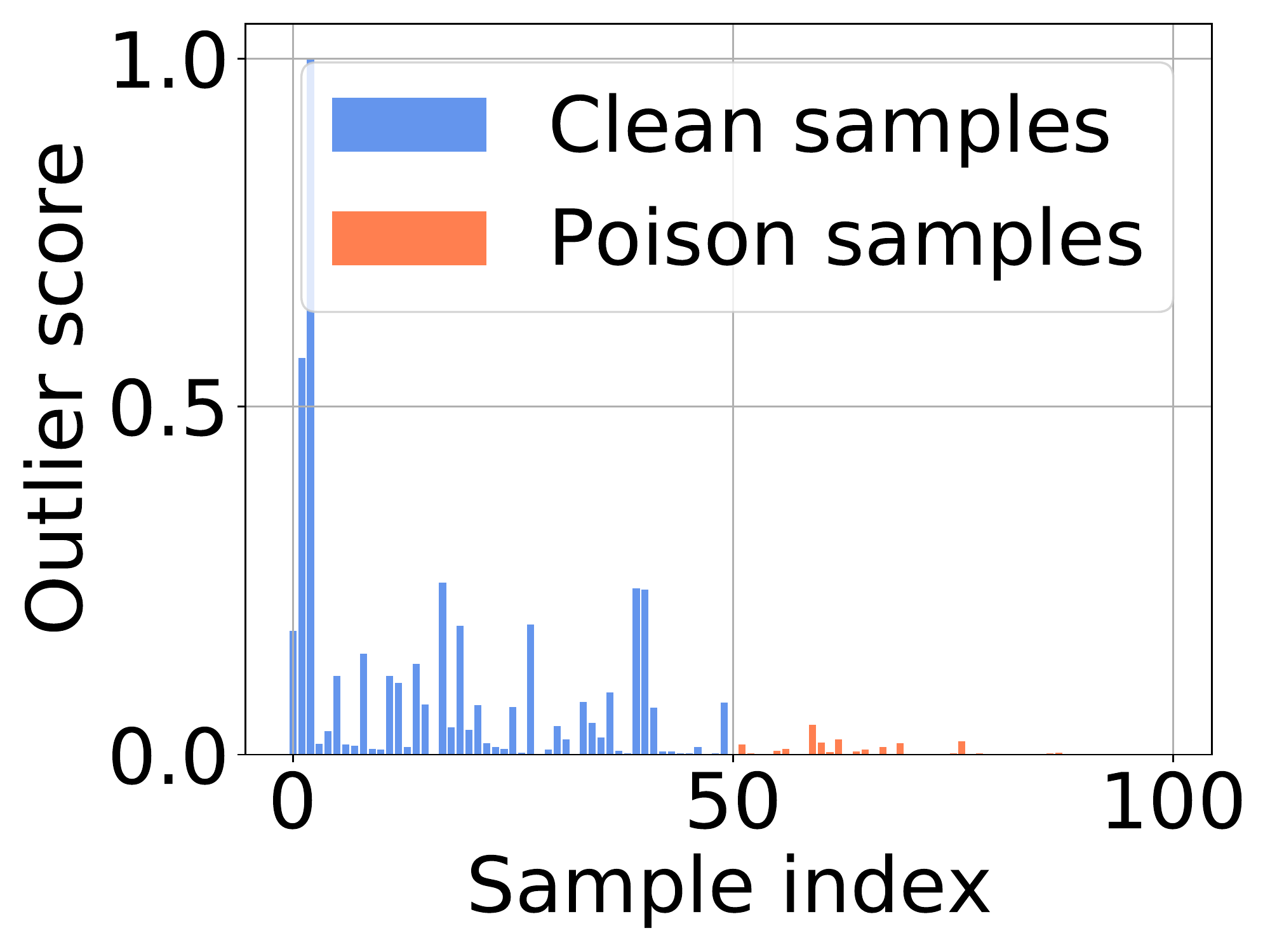}
			\vspace{-0.6cm}
		\end{minipage}
	}
	\caption{\small The outlier score of samples generated by Spectral Signature. The larger the score is, the more likely the sample is an outlier.}
	\label{fig:SpectralSignatures}
	\vspace{-1em}
\end{figure}

\noindent \textbf{Resistance to Spectral Signatures.} Spectral Signatures \cite{NEURIPS2018_280cf18b} discovered the backdoor attacks can leave behind a detectable trace in the spectrum of the covariance of a feature representation. The trace is so-called Spectral Signatures, which is detected using singular value decomposition. This method calculates an outlier score for each sample. It succeeds if clean samples have small values and poison samples have large values (see supplementary material for more details). As shown in Figure \ref{fig:SpectralSignatures}, we test $100$ samples, where $0 \sim 49$ are clean samples and $50 \sim 100$ are poison samples. Our attack notably disturbs this method in that the clean samples have unexpected large scores. 

\vspace{-0.1cm}
\subsection{Discussion}
\vspace{-0.1cm}
In this section, unless otherwise specified, all settings are the same as those stated in Section \ref{sec:settings}.

\vspace{0.2em}
\noindent \textbf{Attack with Different Target Labels. } 
We test our method using different target labels ($y_t = 1,2,3$). Table \ref{tab:targetlabel} shows the BA/ASR of our attack, which reveals the effectiveness of our method using different target labels.

\begin{table}[ht]
	\vspace{-0.4em}
	\caption{\small The BA/ASR (\%) of our attack with other target labels.}
	\scalebox{0.73}{
		\begin{tabular}{cc|cc|cc}
			\hline
			\multicolumn{2}{c|}{Target Label$=1$} & \multicolumn{2}{c|}{Target Label$=2$} & \multicolumn{2}{c}{Target Label$=3$} \\ \hline
			ImageNet          & MS-Celeb          & ImageNet           & MS-Celeb         & ImageNet          & MS-Celeb         \\
			85.4/99.4         & 97.3/99.9         & 85.6/99.3          & 97.6/100         & 85.6/99.5         & 97.2/99.9        \\ \hline
		\end{tabular}
	}
	\vspace{-0.4em}
	\label{tab:targetlabel}
\end{table}

\begin{table}[ht]
\centering
\small
\caption{\small The ASR (\%) of our attack with consistent (dubbed \emph{Ours}) or inconsistent (dubbed \emph{Ours (inconsistent)}) triggers. The inconsistent trigger is generated based on a different testing image.}
\begin{tabular}{c|c|c}
\hline
                                & ImageNet & MS-Celeb-1M \\ \hline
Ours           &   99.5    &     100        \\
Ours (inconsistent) &   23.3      &     98.1        \\ \hline
\end{tabular}
\label{tab:exclusive}
\vspace{-0.3em}
\end{table}

\begin{figure}[ht]
\centering
\includegraphics[width=0.75\linewidth]{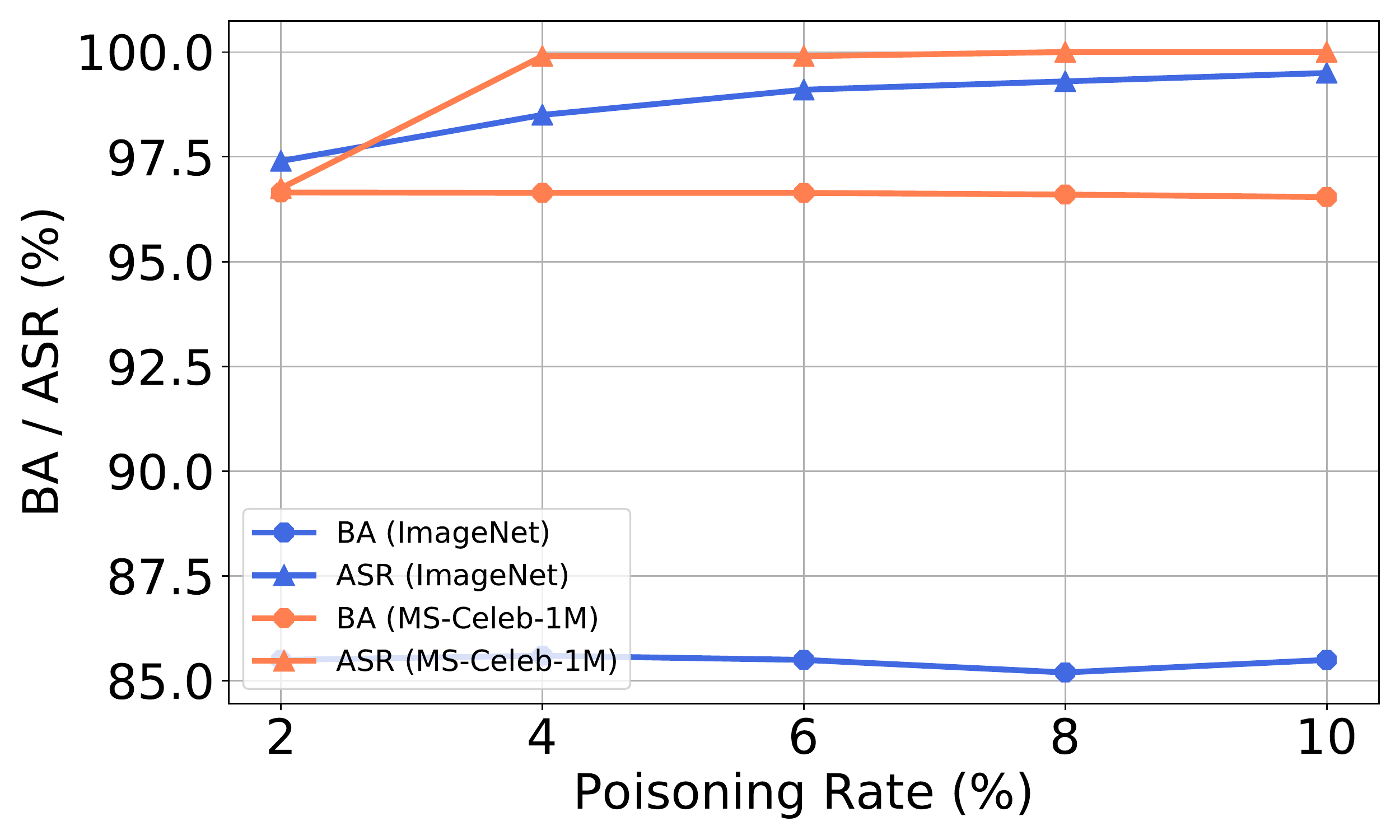}
\vspace{-0.2cm}
\caption{\small The effect of poisoning rate towards our attack.}
\label{fig:poisoning-rate}
\vspace{-0.3em}
\end{figure}

\noindent \textbf{The Effect of Poisoning Rate $\gamma$. } 
In this part, we discuss the effect of the poisoning rate $\gamma$ towards ASR and BA in our attack. As shown in Figure \ref{fig:poisoning-rate}, our attack reaches a high ASR ($>95\%$) on both datasets by poisoning only 2\% of training samples. Besides, the ASR increases with an increase of $\gamma$ while the BA remains almost unchanged. In other words, there is almost no trade-off between the ASR and BA in our method. However, the increase of $\gamma$ will also decrease the attack stealthiness. Attackers need to specify this parameter for their specific needs.

\noindent \textbf{The Exclusiveness of Generated Triggers. }
In this part, we explore whether the generated sample-specific triggers are exclusive, $i.e.,$ whether testing image with trigger generated based on another image can also activate the hidden backdoor of DNNs attacked by our method. Specifically, for each testing image $\bm{x}$, we randomly select another testing image $\bm{x}' \ (\bm{x}' \neq \bm{x})$. Now we query the attacked DNNs with $\bm{x} + T(G(\bm{x}'))$ (rather than with $\bm{x} + T(G(\bm{x}))$). As shown in Table \ref{tab:exclusive}, the ASR decreases sharply when inconsistent triggers ($i.e.$, triggers generated based on different images) are adapted on the ImageNet dataset. However, on the MS-Celeb-1M dataset, attacking with inconsistent triggers can still achieve a high ASR. This may probably be because most of the facial features are similar and therefore the learned trigger has better generalization. We will further explore this interesting phenomenon in our future work.

\begin{table}[t]
\centering
\small
\caption{\small Out-of-dataset generalization of our method in the attack stage. See text for details. }
\vspace{0.1em}
\begin{tabular}{c|cc|cc}
\hline
Dataset for Classifier $\rightarrow$ & \multicolumn{2}{c|}{ImageNet} & \multicolumn{2}{c}{MS-Celeb-1M} \\ \hline
Dataset for Encoder $\downarrow$   & BA            & ASR           & BA             & ASR            \\ \hline
ImageNet               &       85.5        &    99.5           &     95.6           & 99.5               \\
MS-Celeb-1M            &      85.1         &    99.4           &    96.5            &  100              \\ \hline
\end{tabular}
\label{tab:gen_attack}
\vspace{-1em}
\end{table}

\noindent \textbf{Out-of-dataset Generalization in the Attack Stage. }
Recall that the encoder is trained on the benign version of the poisoned training set in previous experiments. In this part, we explore whether the one trained on another dataset can still be adapted for generating poisoned samples of a new dataset (without any fine-tuning) in our attack. As shown in Table \ref{tab:gen_attack}, the effectiveness of attack with an encoder trained on another dataset is on par with that of the one trained on the same dataset. In other words, attackers can reuse already trained encoders to generate poisoned samples, if their image size is the same. \emph{This property will significantly reduce the computational cost of our attack}.

\begin{table}[t]
\centering
\small
\caption{\small The ASR (\%) of our method attacked with out-of-dataset testing samples. See text for details.}
\begin{tabular}{c|c|c}
\hline
\tabincell{c}{Dataset for Training $\rightarrow$ \\ Dataset for Inference $\downarrow$} & ImageNet & MS-Celeb-1M \\ \hline
Microsoft COCO                  &    100      &     99.9        \\
Random Noise          &    100      &       99.9      \\ \hline
\end{tabular}
\label{tab:gen_inference}
\vspace{-1em}
\end{table}

\noindent \textbf{Out-of-dataset Generalization in the Inference Stage. }
In this part, we verify that whether out-of-dataset images (with triggers) can successfully attack DNNs attacked by our method. We select the Microsoft COCO dataset \cite{lin2014microsoft} and a synthetic noise dataset for the experiment. They are representative of nature images and synthetic images, respectively. Specifically, we randomly select 1,000 images from the Microsoft COCO and generate 1,000 synthetic images where each pixel value is uniformly and randomly selected from $\{0, \cdots, 255\}$. All selected images are resized to $3 \times 224 \times 224$. As shown in Table \ref{tab:gen_inference}, our attack with poisoned samples generated based on out-of-dataset images can also achieve nearly 100\% ASR. \emph{It indicates that attackers can activate the hidden backdoor in attacked DNNs with out-of-dataset images (not necessary with testing images)}.

\comment{
\vspace{0.3em}
\noindent \textbf{Connection Between the Failure of Attack and of Decoder. }

\red{I expect that the ASR of samples whose contained string can be successfully decoded will be much higher than those that can not be successfully decoded.}

\begin{table}[ht]
\centering
\small
\caption{\small The BA (\%) and ASR (\%) of testing images whose contained string can be successfully and unsuccessfully decoded by the decoder. }
\vspace{0.1em}
\scalebox{0.95}{
\begin{tabular}{c|cc|cc}
\hline
Dataset $\rightarrow$ & \multicolumn{2}{c|}{ImageNet} & \multicolumn{2}{c}{MS-Celeb-1M} \\ \hline
Image Type $\downarrow$   & BA            & ASR           & BA             & ASR            \\ \hline
\lyz{With Decoder}               &       85.5        &      99.5         &       96.5         &        100        \\
\lyz{Without Decoder}              &      85.2       &               &            96.7    &         98       \\ \hline
\end{tabular}
}
\end{table}
}

\vspace{-0.3cm}
\section{Conclusion}
\vspace{-0.1cm}
In this paper, we showed that existing backdoor attacks were easily alleviated by current backdoor defenses mostly because their backdoor trigger is sample-agnostic, $i.e.$, different poisoned samples contain the same trigger. Based on this understanding, we explored a new attack paradigm, the sample-specific backdoor attack (SSBA), where the backdoor trigger is sample-specific. Our attack broke the fundamental assumption of defenses, therefore can bypass them. Specifically, we generated sample-specific invisible additive noises as backdoor triggers by encoding an attacker-specified string into benign images, motivated by the DNN-based image steganography. The mapping from the string to the target label will be learned when DNNs are trained on the poisoned dataset. Extensive experiments were conducted, which verify the effectiveness of our method in attacking models with or without defenses. 

\vspace{0.2em}
\noindent \textbf{Acknowledgment.} 
Yuezun Li is supported in part by China Postdoc Science Foundation under grant No.2021TQ0314.
Baoyuan Wu is supported by the Natural Science
Foundation of China under grant No.62076213, the university development fund of the Chinese University of Hong Kong, Shenzhen under grant No.01001810, the special project fund of Shenzhen Research Institute of Big Data under grant No.T00120210003, and Shenzhen Science and Technology Program under grant No.GXWD20201231105722002-20200901175001001. Siwei Lyu is supported by the Natural Science
Foundation under grants No.IIS-2103450 and IIS-1816227. 

\newpage

{\small
\bibliographystyle{ieee_fullname}
\bibliography{egbib}
}

\clearpage

\setcounter{section}{0}
\section*{Appendix}

\begin{table}[ht]
\centering
\small
\caption{The BA (\%) and ASR (\%) of methods with VGG-16. Among all attacks, the best result is denoted in boldface while underline indicates the second-best result.}
\begin{tabular}{c|cc|cc}
\hline
Dataset $\rightarrow$           & \multicolumn{2}{c|}{ImageNet} & \multicolumn{2}{c}{MS-Celeb-1M} \\ \hline
Attack $\downarrow$, Metric $\rightarrow$    & BA            & ASR           & BA             & ASR            \\ \hline
Standard Training & 83.9              &  0             &  96.9              &   0.1             \\ \hline
BadNets           & $\bm{84.6}$          & $\bm{100}$           & $\underline{95.8}$           & $\bm{100}$            \\
Blended Attack    & $\underline{84.3}$          & 96.9          & 95.5           & $\underline{99.2}$           \\
Ours              & 83.5          & $\underline{98.6}$          & $\bm{96.3}$           & $\bm{100}$            \\ \hline
\end{tabular}
\label{tab:main_vgg}
\vspace{-1em}
\end{table}

\section{More Results of Methods with VGG-16}
In the main manuscript, we used ResNet-18 \cite{he2016deep} as the model structure for all experiments. To verify that our proposed attack is also effective towards other model structures, we provide additional results of methods with VGG-16 \cite{simonyan2014very} in this section. Unless otherwise specified, all settings are the same as those used in the main manuscript.

\subsection{Attack Effectiveness}
Follow the settings adopted in the main manuscript, we compare the effectiveness of methods from the aspect of attack success rate (ASR) and benign accuracy (BA).

As shown in Table \ref{tab:main_vgg}, our attack can also reach a high attack success rate and benign accuracy on both ImageNet and MS-Celeb-1M dataset with VGG-16 as the model structure. Specifically, our attack can achieve an ASR $>98.5\%$ on both datasets. Moreover, the ASR of our attack is on par with that of BadNets and higher than that of the Blended Attack. These results verify that sample-specific invisible additive noises can also serve as good backdoor triggers even though they are more complicated than the white-square used in BadNets and Blended Attack.

\begin{figure}[ht]
    \centering
    \includegraphics[width=0.9\linewidth]{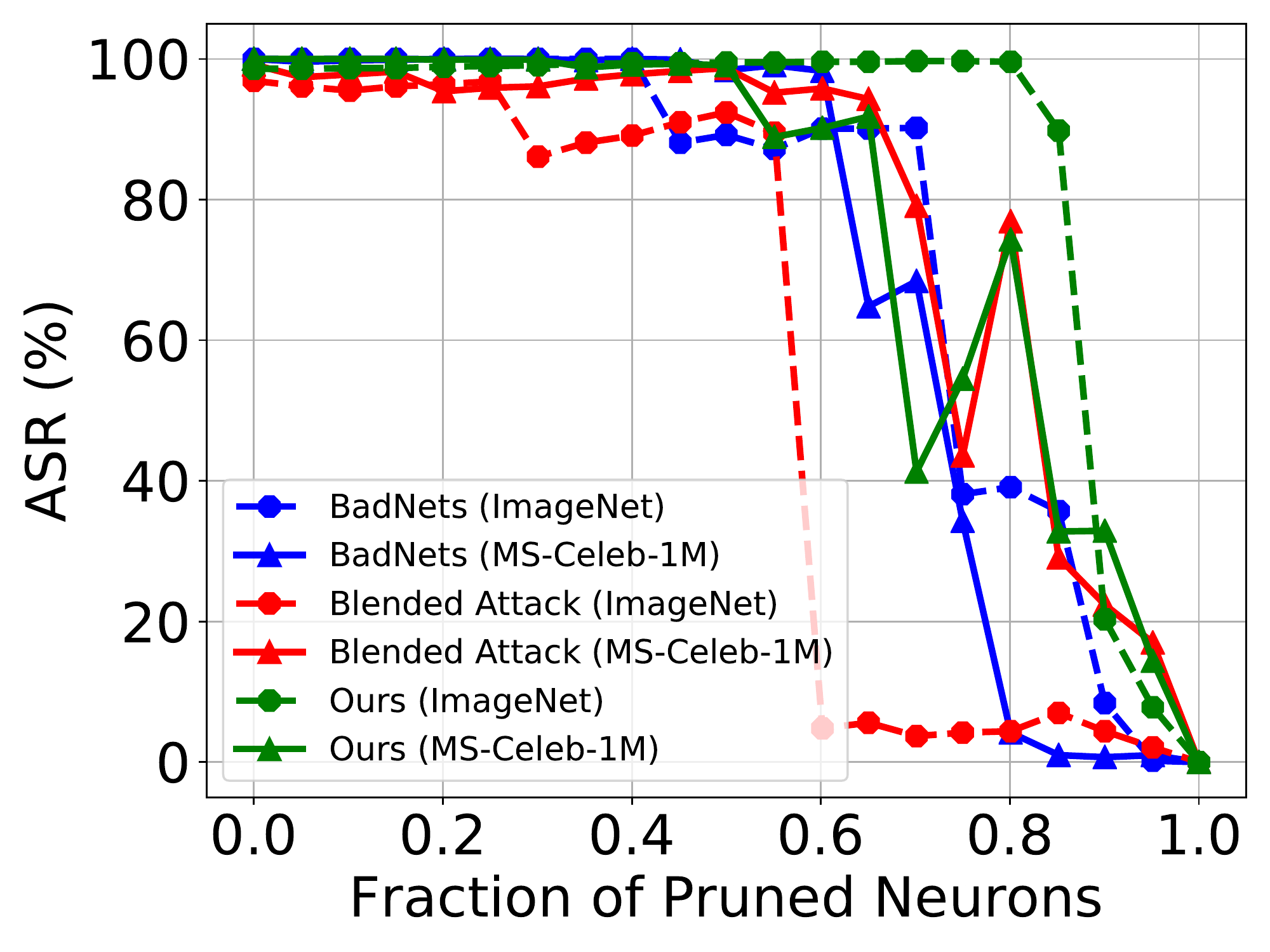}
    \caption{\small The ASR (\%) of different attacks $w.r.t.$ the fraction of pruned neurons on the ImageNet and MS-Celeb-1M dataset.}
    \label{fig:vgg-prune}
\end{figure}

\subsection{Resistance to Fine-Pruning}
In this part, we also compare our attack with the BadNets and Blended Attack in terms of the resistance to the pruning-based defense \cite{liu2018fine}. As shown in Figure \ref{fig:vgg-prune}, curves of our attack are always above those of other attacks. In other words, our descent speed is slower although ASRs of all attacks decrease with the increase of the fraction of pruned neurons. For example, on the ImageNet dataset, the ASR of Blended Attack decrease to less than 10\% when 60\% neurons are pruned, whereas our attack still preserves a high ASR ($>95\%$). This suggests that our attack is more resistant to the pruning-based defense.

\subsection{Resistance to Neural Cleanse}
In this part, we also compare our attack with the BadNets and Blended Attack in terms of the resistance to the Neural Cleanse \cite{wang2019neural}. Recall that there are two indispensable requirements for the success of Neural Cleanse, including \textbf{(1)} successful select one candidate ($i.e.$, the anomaly index is big enough) and \textbf{(2)} the selected candidate is close to the backdoor trigger.

\begin{figure*}[ht]
    \centering
    \includegraphics[width=\linewidth]{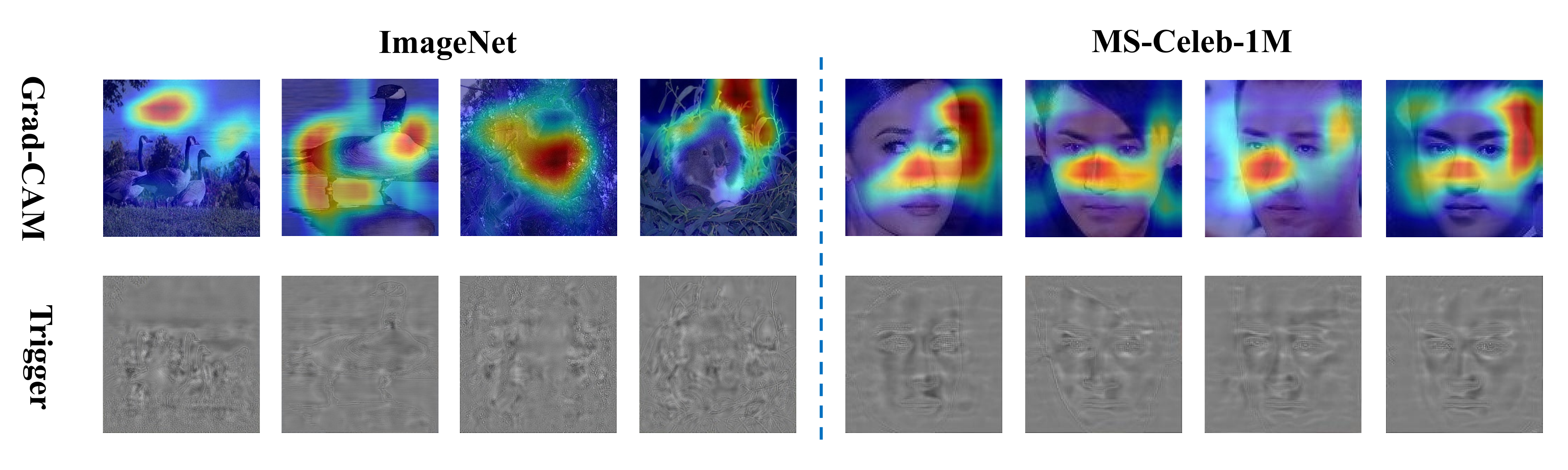}
    \vspace{-0.5em}
    \caption{\small The Grad-CAM of poisoned samples and their corresponding triggers of our attack.}
    \label{fig:cam_vgg}
    \vspace{-0.8em}
\end{figure*}

\begin{figure}[ht]
\centering
\subfigure[ImageNet]{
\begin{minipage}[b]{0.473\linewidth}
\centering
\includegraphics[width=\textwidth]{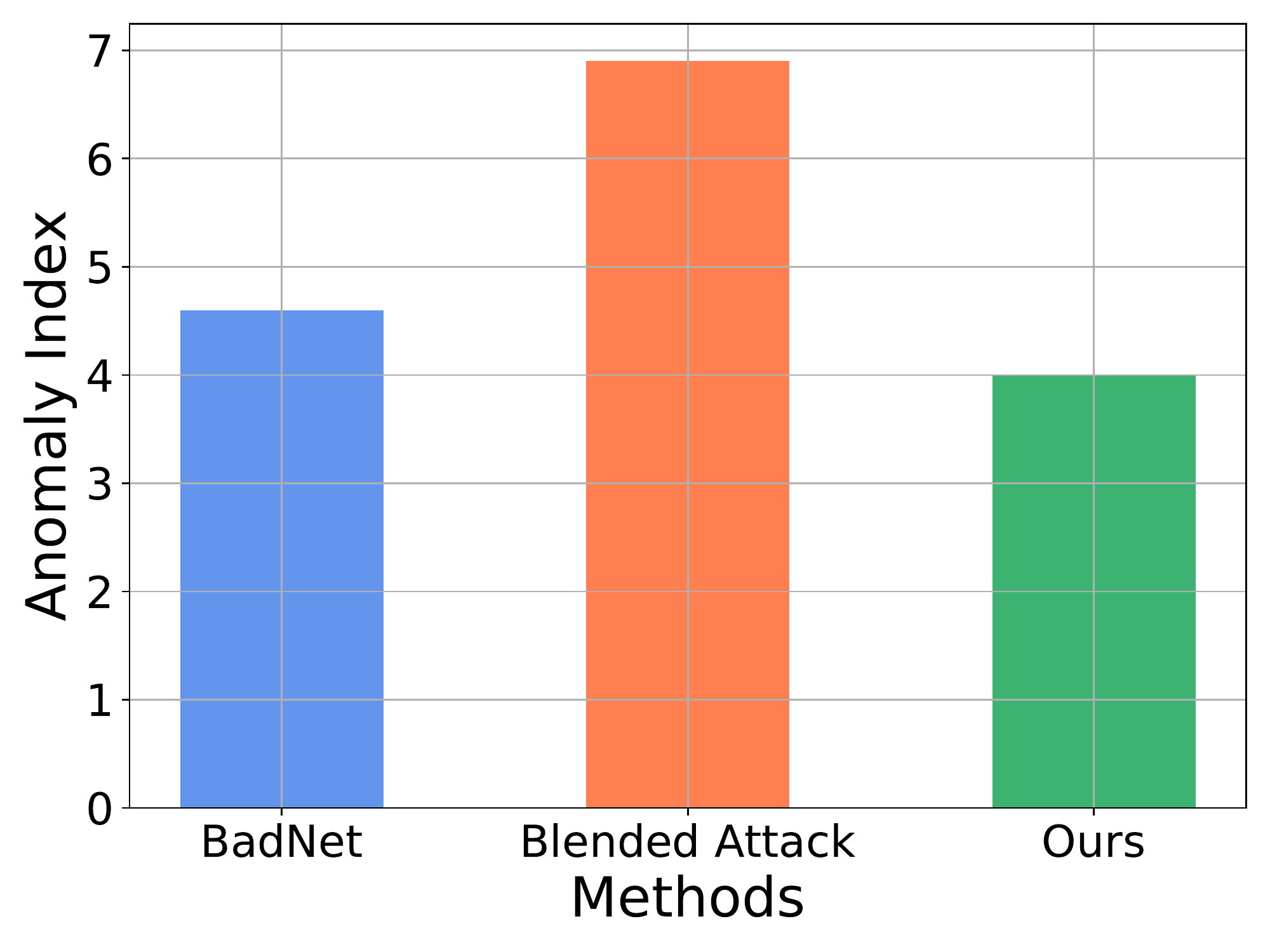}
\end{minipage}
}
\subfigure[MS-Celeb-1M]{
\begin{minipage}[b]{0.473\linewidth}
\centering
\includegraphics[width=\textwidth]{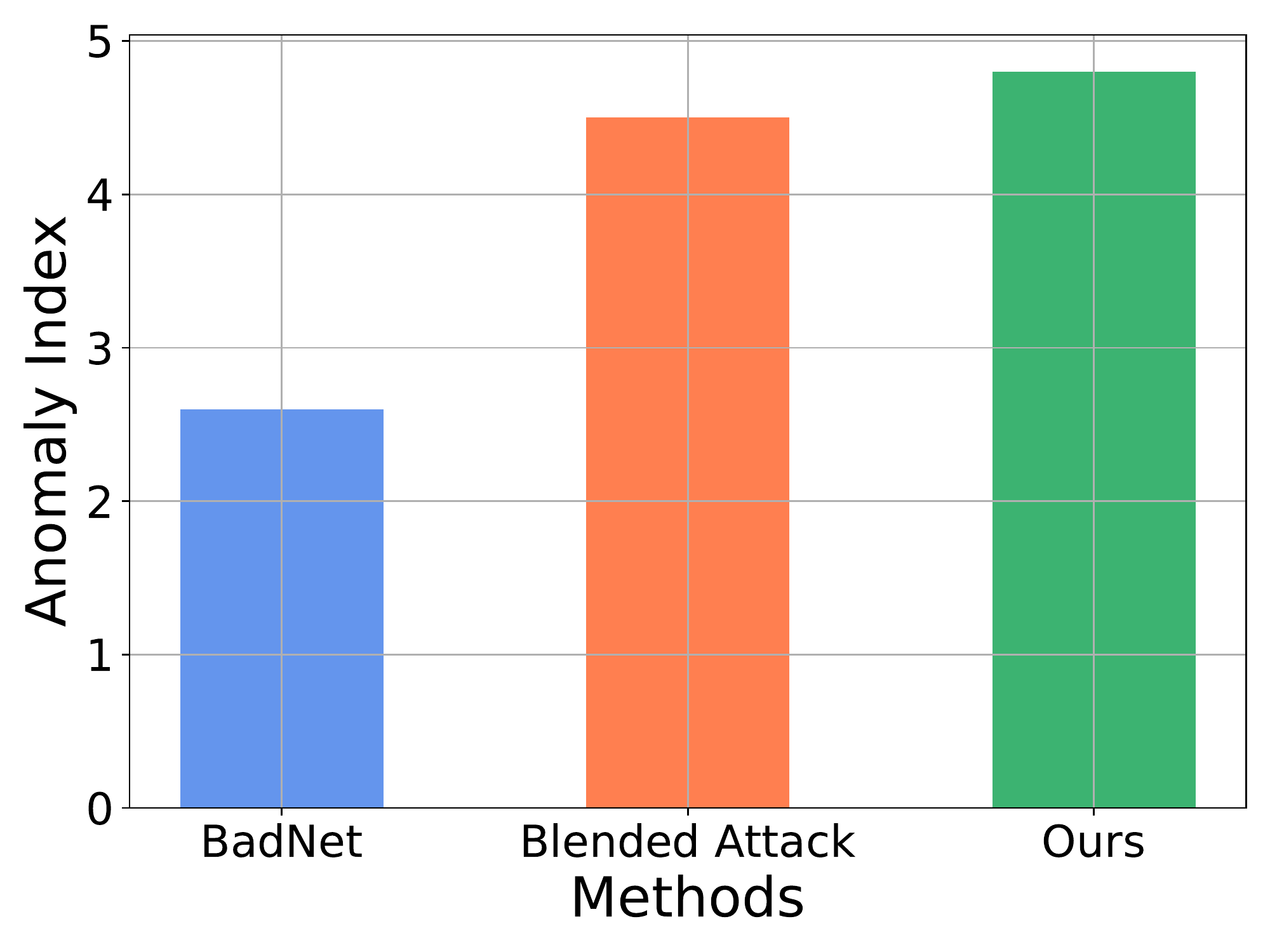}
\end{minipage}
}
\caption{\small The anomaly index of different attacks with VGG-16 on the ImageNet and MS-Celeb-1M dataset. The smaller the index, the harder the attack for Neural-Cleanse to defend.}
\label{fig:neural-cleanse_vgg}
\vspace{-0.3em}
\end{figure}

\begin{figure}[ht]
\centering
\includegraphics[width=\linewidth]{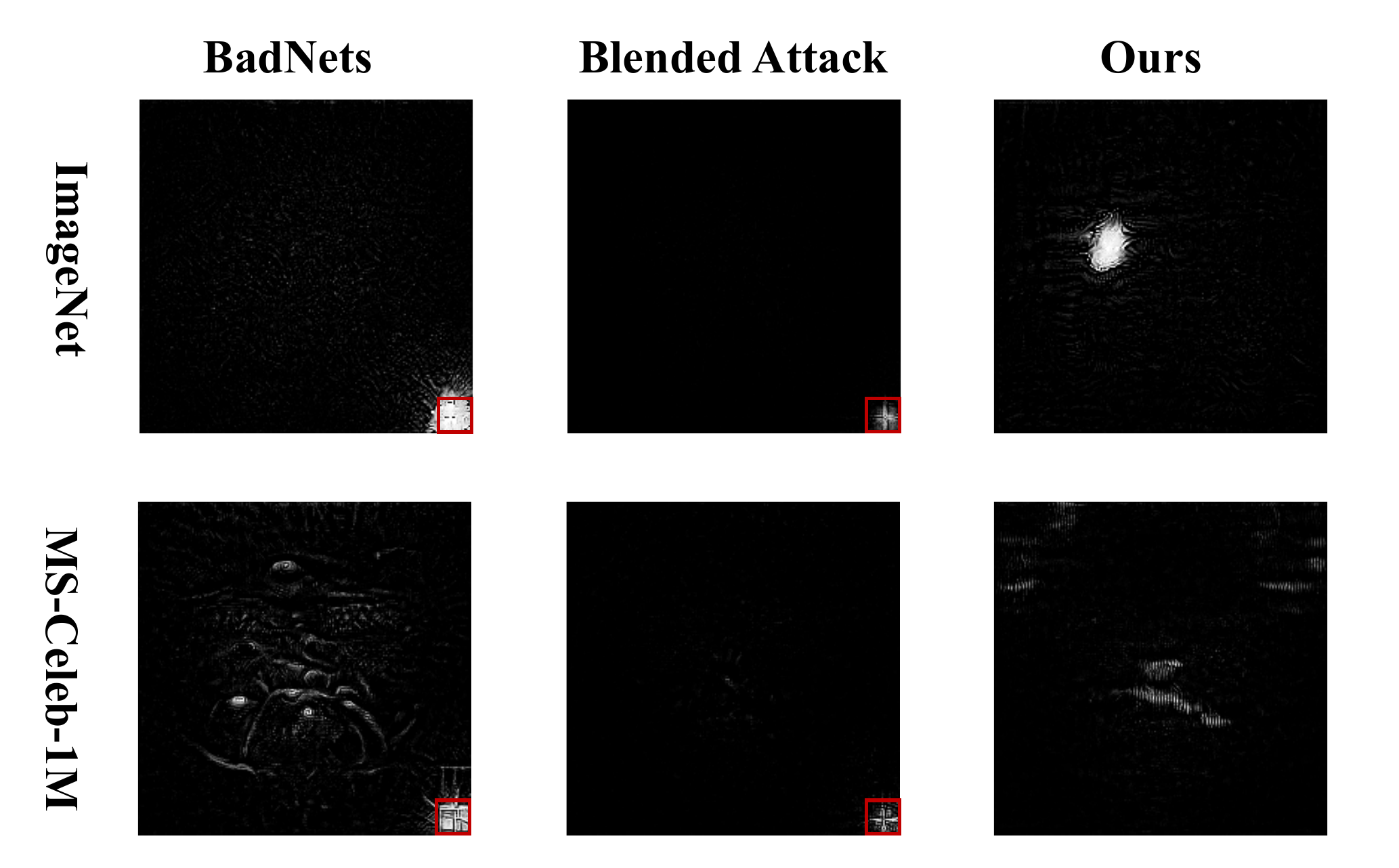}
\caption{\small The synthesized triggers generated by Neural Cleanse. Red box in the figure indicates ground-truth trigger areas.}
\label{fig:sys_vgg}

\end{figure}

As shown in Figure \ref{fig:neural-cleanse_vgg}, the anomaly index
of our attack is smaller than that of BadNets and Blended Attack on the ImageNet dataset. In other words, our attack is more resistant to the Neural Cleanse in this case. We also visualize the synthesized trigger ($i.e.$, the one with the smallest anomaly index among all candidates) of different attacks. As shown in Figure \ref{fig:sys_vgg}, although our attack reaches the highest anomaly index on the MS-Celeb-1M dataset, synthesized triggers of our attack are meaningless. In contrast, synthesized triggers of BadNets and Blended Attack contain similar patterns to the ones used by attackers. As such, our attack is still more resistant to the Neural Cleanse in this case.

\begin{figure}[ht]
\centering
\subfigure[ImageNet]{
\begin{minipage}[b]{0.473\linewidth}
\centering
\includegraphics[width=\textwidth]{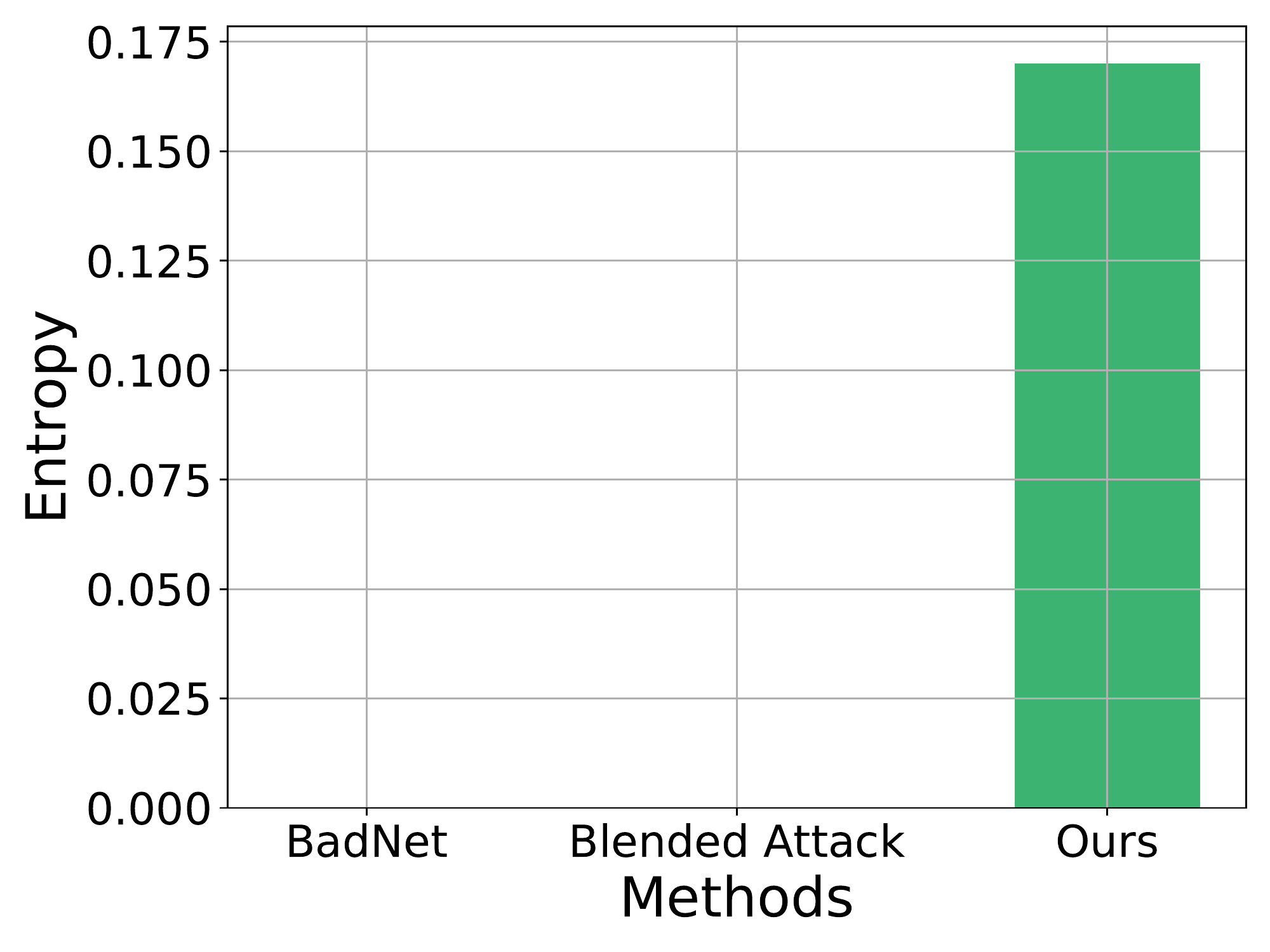}
\end{minipage}
}
\subfigure[MS-Celeb-1M]{
\begin{minipage}[b]{0.473\linewidth}
\centering
\includegraphics[width=\textwidth]{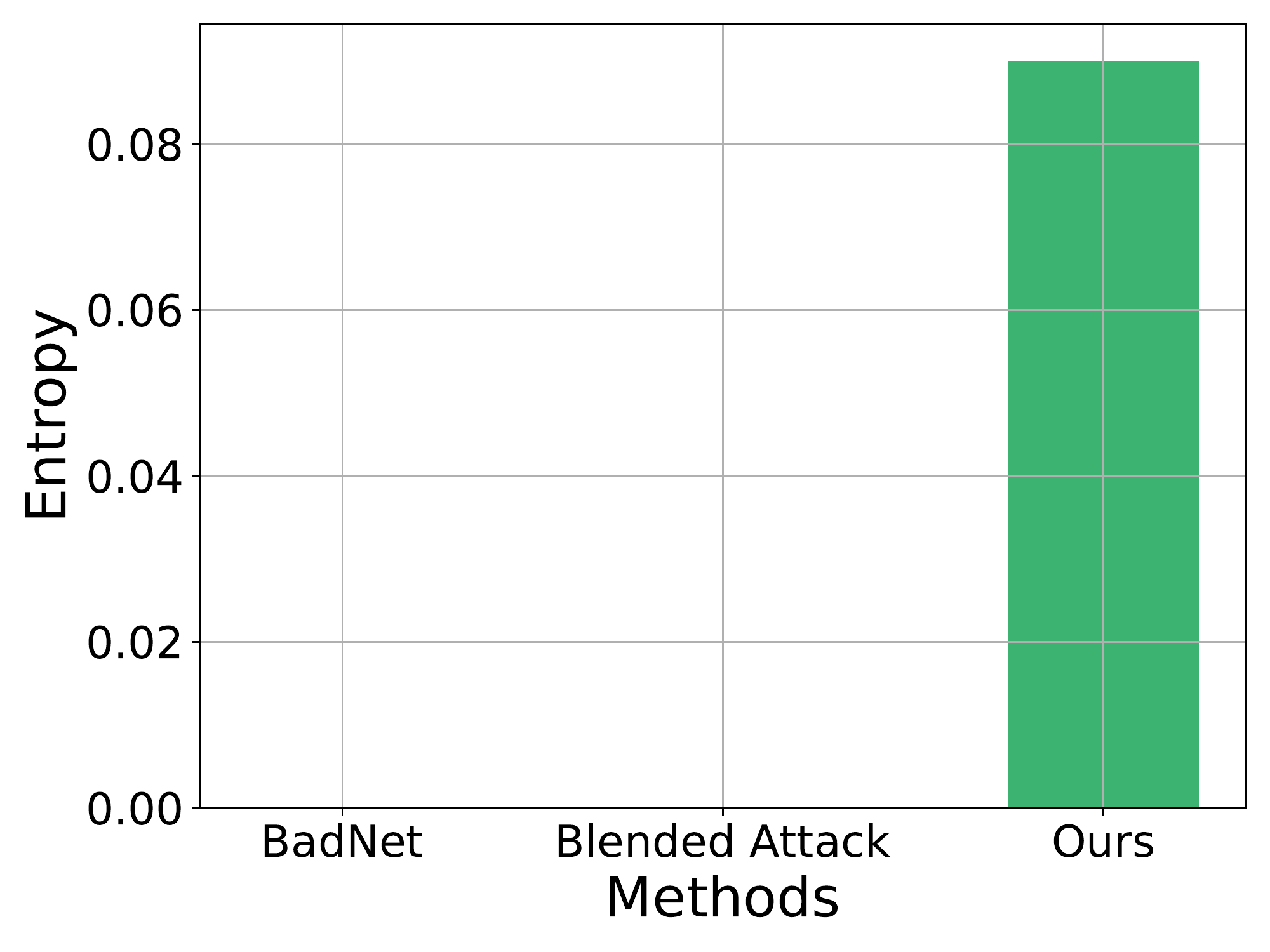}
\end{minipage}
}
\caption{\small The entropy generated by STRIP of different attacks. The higher the entropy, the harder the attack for STRIP to defend.}
\label{fig:strip_vgg}
\end{figure}

\subsection{Resistance to STRIP}

STRIP \cite{gao2019strip} filters poisoned samples based on the prediction randomness of samples generated by imposing various image patterns on the suspicious image. The randomness is measured by the entropy of the average prediction of those samples. As such, the higher the entropy, the harder an attack for STRIP to defend. As shown in Figure \ref{fig:strip_vgg}, our attack has a significantly higher entropy compared with other baseline methods on both ImageNet and MS-Celeb-1M datasets. In other words, our attack is more resistant to the STRIP compared with other attacks.

\subsection{Resistance to SentiNet}
SentiNet \cite{chou2020sentinet} identities trigger regions based on the similarities of Grad-CAM of different samples. As shown in Figure \ref{fig:cam_vgg}, Grad-CAM fails to detect trigger regions of images generated by our attack. Besides, the Grad-CAM of different poisoned samples has a significant difference. As such, our attack can bypass the SentiNet.

\section{Detailed Settings of DF-TND and Spectral Signature}

\noindent{\bf DF-TND.} Note the vanilla setting of DF-TND is selected based on the CIFAR dataset, rather than the datasets used in our experiment. We found that its performance is sensitive to the hyper-parameter values. To achieve a fairer comparison, we fine-tune their hyper-parameters to seek a best-performed setting, based on the criteria that the more front of target label in a descending order based on logit increase denotes better defensive performance. We fine-tune two hyper-parameters, which are the batch size $b$ of testing random noise images and the sparsity parameter $\gamma$ used in the adversarial attack. In its vanilla setting, the batch size $b$ is set to $10$ and $\gamma$ is set to $0.001$. In our experiments, we test nine hyper-parameter combinations, where batch size $b$ is selected from $\{10,20,30\}$ and sparsity parameter $\gamma$ is selected from $\{0.00001,0.0001,0.001\}$ and then select the best-performed hyper-parameter combination. Specifically, we select $b=10,\gamma=0.0001$ for ImageNet dataset and $b=20,\gamma=0.00001$ for MS-Celeb-1M dataset.

\smallskip
\noindent{\bf Spectral Signature.} Since this work does not release the code, we implement it based on Trojan-Zoo\footnote{\url{https://github.com/alps-lab/Trojan-Zoo}}. Similar to DF-TND, Spectral Signature is also designed for CIFAR dataset, such that the default threshold of outlier score is not applicable in our experiments. For fair comparison, we calculate the outlier score for each test sample and show the distribution instead. The defense fails if the clean samples have larger outlier scores.

\section{More Comparisons with Adapted Methods}
\begin{figure}[ht]
	\centering
	\raisebox{2em}{
		\begin{turn}{90}
			\cite{saha2019hidden}
		\end{turn}
	}
	\includegraphics[width=0.25\linewidth]{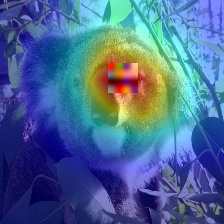} 
	\includegraphics[width=0.25\linewidth]{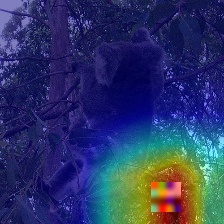} 
	\includegraphics[width=0.25\linewidth]{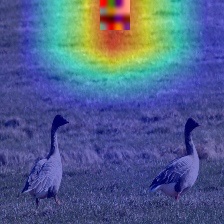} \\	
	\raisebox{2em}{
		\begin{turn}{90}
			\cite{zhao2020clean}
		\end{turn}
	}
	\includegraphics[width=0.25\linewidth]{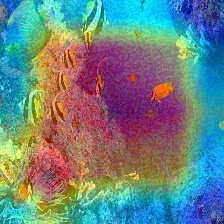} 
	\includegraphics[width=0.25\linewidth]{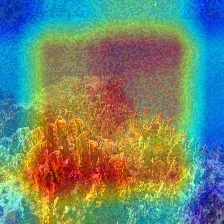} 
	\includegraphics[width=0.25\linewidth]{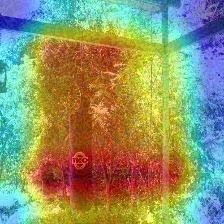} \\
	\caption{\small The Grad-CAM of poisoned samples generated by different methods.}
    \label{fig:adapted-cam}
\end{figure}

As aforementioned in Section 2, the works \cite{quiring2020backdooring,saha2019hidden,zhao2020clean} are out of our scope either in the task or threat model. However, to be more comprehensive, we attempt to adapt the code of \cite{saha2019hidden,zhao2020clean} to our scenario for comparison. Note \cite{saha2019hidden} and \cite{zhao2020clean} are originally validated with AlexNet and CNN+LSTM respectively. We change their backbones to ResNet-18 and abandon their clean-label setting for fair comparison. The triggers of \cite{saha2019hidden} and \cite{zhao2020clean} are movable specific block and targeted universal adversarial perturbation (UAP) respectively. Table \ref{tab:adapted} shows the BA/ASR on ImageNet without defense, which represents our adaptations of these methods are normal. Figure \ref{fig:adapted-cam} shows the Grad-CAM of SentiNet defense, where the block trigger of \cite{saha2019hidden} is accurately localized and the UAP trigger of \cite{zhao2020clean} is stably identified.

\begin{table}[ht]
	\centering
    \caption{\small The BA/ASR (\%) performance of ResNet-18 on ImageNet dataset.}
	\begin{tabular}{c|c}
		\hline
		Methods & BA/ASR \\ \hline
		\cite{saha2019hidden}            &   84.4/99.8  \\\hline
		\cite{zhao2020clean}            &   85.5/99.9  \\ \hline
		Ours            &   85.5/99.5  \\ \hline
	\end{tabular}
	\label{tab:adapted}
\end{table}

\end{document}